\documentclass[prb,twocolumn,aps,floatfix,longbibliography]{revtex4-2}
\usepackage{graphicx,amsmath,xcolor}
\usepackage{hyperref}
\usepackage{subcaption}
\usepackage{float}
\usepackage[normalem]{ulem}
\usepackage{soul}
\usepackage{silence}
\usepackage{mathtools}
\usepackage[T1]{fontenc}
\DeclarePairedDelimiterX\innerp[2]{\langle}{\rangle}{#1\delimsize\vert\mathopen{}#2}%
\DeclarePairedDelimiterX\braket[2]{\langle}{\rangle}{#1\delimsize\vert\mathopen{}#2}%
\DeclarePairedDelimiterX\braketOP[3]{\langle}{\rangle}{#1\,\delimsize\vert\,\mathopen{}#2\,\delimsize\vert\,\mathopen{}#3}%
\DeclarePairedDelimiterX\ketbra[2]{\lvert}{\rvert}{#1\delimsize\rangle\!\delimsize\langle#2}%
\DeclarePairedDelimiterX\outerp[2]{\lvert}{\rvert}{#1\delimsize\rangle\!\delimsize\langle#2}%
\DeclarePairedDelimiterX\projector[1]{\lvert}{\rvert}{#1\delimsize\rangle\!\delimsize\langle#1}%

\newcommand{\comment}[1]{}

\begin{document}
    \title{Neural network based deep learning analysis of semiconductor quantum dot qubits for automated control}
    \author{Jacob R. Taylor}
    \author{Sankar Das Sarma}
    \affiliation{Condensed Matter Theory Center and Joint Quantum Institute, Department of Physics, University of Maryland, College Park, Maryland 20742-4111 USA}
\begin{abstract}

{\color{purple} \comment{Machine learning offers a largely unexplored avenue for improving noisy disordered devices in physics using automated algorithms. Through simulations that include disorder in physical devices, particularly quantum devices, there is potential to learn about disordered landscapes and subsequently tune devices based on those insights.}} In this work, we introduce a novel methodology that employs machine learning, specifically convolutional neural networks (CNNs), to discern the disorder landscape in the parameters of the disordered extended Hubbard model underlying the semiconductor quantum dot spin qubit architectures. { Our approach, which we demonstrate with numerically simulated devices} takes advantage of { currently used} experimentally obtainable charge stability diagrams from neighboring quantum dot pairs, enabling the CNN to accurately identify disorder in each parameter of the extended Hubbard model. Remarkably, our CNN can process site-specific disorder in Hubbard parameters, including variations in hopping constants, on-site potentials (gate voltages), and both intra-site and inter-site Coulomb terms. This advancement facilitates the prediction of spatially dependent disorder across all parameters simultaneously with high accuracy ($R^2>0.994$) and fewer parameter constraints, marking a significant improvement over previous methods that were focused only on analyzing on-site potentials at low coupling. Furthermore, our approach allows for the tuning of five or more quantum dots at a time, effectively addressing the often-overlooked issue of crosstalk. Not only does our method streamline the tuning process, potentially enabling fully automated adjustments, but it also introduces a "no trust" verification method to rigorously validate the neural network's predictions. Ultimately, this work aims to lay the groundwork for generalizing our method to tackle a broad spectrum of physical problems. {\comment{In particular, our work establishes that the microscopic parameters controlling the semiconductor quantum dot quantum computing platforms can be uniquely determined in an automated manner by using a CNN based machine learning technique using only the measured charge stability diagrams as the input.}}

\end{abstract}
\maketitle 

Two current areas of high scientific and technological (in fact, even public) interest are neural networks and quantum computation. While neural networks {\comment{(or more generally, machine learning)}} have progressed much more rapidly recently, quantum computation has also been making significant progress. Machine learning provides a tool that could potentially speed up advances in the development of quantum information hardware. The issue of interest to us is the inevitable presence of unknown and unintentional random disorder in the quantum devices, which necessitates repeated finetuning and re-calibration of the devices in order to obtain meaningful and reproducible experimental results.  We propose that neural network based machine learning could provide considerable help in this tuning problem in an unknown disorder landscape. In particular, for a large class of physical problems, there exists a sufficiently accurate physical model that is disordered in an unknown but potentially predictable manner. For this class of problems, it may be possible through theoretical simulations to generate enough training data that allows a neural network to determine the disorder landscape of the problem and then feasibly correct it. 

One case we will examine here is the disorder within quantum dots as occurring in the important (because it is scalable) semiconductor quantum dot based spin qubit platforms \cite{hanson2007spins,sarma2001spin,van2002electron,sarma2005spin,chatterjee2021semiconductor,burkard2023semiconductor}. Quantum dots serve as a promising platform for qubits \cite{hanson2007spins,sarma2001spin,van2002electron,sarma2005spin,chatterjee2021semiconductor,burkard2023semiconductor}. Extensive work has gone into investigating quantum dots for quantum computation, both experimentally and theoretically \cite{hanson2007spins,sarma2001spin,van2002electron,sarma2005spin,chatterjee2021semiconductor,burkard2023semiconductor,loss1998quantum,byrnes2008quantum,zwanenburg2013silicon,zhang2019semiconductor,kobayashi2021engineering,chatterjee2021semiconductor,eng2015isotopically,taylor2023assessing,hensgens2017quantum,hu2000hilbert}. Quantum dots in semiconductors, used as spin qubits have many advantages, such as their long coherence times \cite{kobayashi2021engineering}, especially when isotopically purified \cite{eng2015isotopically}, electric field controllability \cite{reed2016reduced}, and their fast and easy readout \cite{harvey2018high}. But these quantum dots invariably have unknown disorder (arising from random impurities and defects in the system) in the environment, which complicate qubit operations. Our work has the ambitious goal of using machine learning to figure out the disorder landscape so that automated control of qubit operations may become feasible in the future.

The semiconductor microelectronics industry is mature, because of its key role in all modern technologies, thereby enabling the integration of quantum dot technology into the pre-existing infrastructure. Additionally, quantum dots can be created sufficiently small in size consistent with the existing microelectronics industry, allowing for scalability to large number of qubits required to perform advanced quantum algorithms \cite{shor1994algorithms}. This is a huge potential scalability and manufacturing advantage driving quantum computing research in the quantum dot based qubit platforms. In fact, recently Si quantum dot qubits have been fabricated in advanced manufacturing fabs where industrial semiconductor wafers are made, establishing the microelectronic manufacturing compatibility and potential of quantum dot architectures \cite{neyens2024probing,zwerver2022qubits}. Semiconductor quantum dots in their quantum computing architecture, where the two-qubit gates are operated by inter-dot exchange coupling, are naturally described by the Hubbard model in the low-temperature, strong-interaction regime, which facilitates their numerical simulation \cite{stafford1994collective,kotlyar1998correlated,hensgens2017quantum,sarma2011hubbard,salfi2016quantum}. This is typically set up as a Hubbard model in the form of a chain (or plaquette) of sites {\comment{(with each quantum dot being a site)}}of some length. Disorder significantly impedes the use of quantum dots for computational purposes, making the control of disorder one of the most important challenges to overcome in utilizing quantum dots as a platform for quantum computation. The disorder within quantum dots often manifests as errors in each Hubbard model parameter, which are generally unknown, vary between sites, and must be corrected before utilizing the dots for computation. Automating the control of such disordered quantum dot structures in order to carry out precise gate operations is a serious practical problem since manual control can only work for a system of a few dots \cite{zwolak2023data,zwolak2023colloquium}. Scaling up to a many-dot quantum computing platform requires a decisive resolution of the tuning problem in the form of an automated procedure.

In previous work \cite{hensgens2017quantum}, a method was developed to determine site-dependent disorder in the gate voltage for three quantum dots, using a essentially a by-hand trial and error algorithm to correct the disorder in the gate voltages of the dot chain. This methodology has been invaluable for experimentalists aiming to employ quantum dots for quantum computational purposes. Specifically, this approach utilizes charge stability diagrams, which are easily produced and serve as the primary means for understanding the properties of these devices. { There have also been other works making use of machine learning for this platform \cite{durrer2020automated,wang2023automated,schuff2024fully}. We mention that the main point of our work, a theoretical machine learning based automation of controlling charge stability in semiconductor quantum dot qubits, has recently been identified as one of the central problems preventing progress in semiconductor based solid state qubits \cite{zwolak2023data,zwolak2023colloquium}.}  For quantum dots to be viable for quantum computation, it is crucial that their Hubbard parameters are tuned to specific values so that the Hamiltonian is precisely defined (which is an essential condition for error-free initialization and gate operations). However, this becomes a challenging task as these parameters are prone to natural drift over time due to environmental factors, noise and disorder, necessitating periodic re-tuning through human intervention. The laborious process of re-tuning quantum dots is both unwelcome and time-consuming, typically requiring the examination of numerous charge stability diagrams by-hand to identify the optimal corrections for parameter drift. Current methodologies \cite{hader2024simulation} also limit adjustments to only two quantum dots at a time, since attempting to calibrate all the Hubbard model parameters simultaneously for a sequence of quantum dots is impractical for an individual. Clearly, a much better automatic control must be developed for scaling up the quantum dot structures since calibration and re-tuning individually is out of question for more than 2-3 dots.

Machine learning presents an opportunity for improvement. Neural networks, through the use of large amounts of data, have been able to make connections and thus predictions rapidly for tasks that would require impossibly extensive human effort (e.g. image processing). In physics, machine learning techniques have been utilized to provide valuable insights into disordered problems, including potentially determining the entire disorder landscape \cite{taylor2023machine}. When a sufficiently accurate theoretical model exists {  \comment{(i.e. a precise Hamiltonian such as the Hubbard model for quantum dot qubits)}}- even if it cannot be directly applied due to disorder within its parameters, which take an understandable {\comment{(i.e. not arbitrary)}} but unknown {\comment{(and random)}} form in the Hamiltonian—it is possible to generate large amounts of training data from simulations. This training data can often enable a deep neural network to find the disordered parameters in a way that is applicable to experimental devices. Furthermore, if the device can be modeled accurately, one can conduct verification tests beyond the neural network's input to validate its predictions.

In the current work, we present a neural network based machine learning method that, through a series of experimentally achievable measurements, is able to accurately identify disordered deviations in the parameters of the extended Hubbard model with no prior information about the random disorder incorporated in the theory implicitly or explicitly. It is crucial to emphasize that we are addressing a many-body disordered strong correlation problem, namely the disordered Hubbard model, solely through machine learning. Specifically, we demonstrate that by inputting as the training data a series of charge stability diagrams from nearest neighbor pairs of quantum dots into a convolutional neural network (CNN), it is possible to pinpoint the disorder deviations in each parameter of the extended Hubbard model. The fact that this can be done is prima facie non-obvious, and our results are validated aposteriori (which is of course true for all machine learning and artificial intelligence algorithm-- the proof of its efficacy is based on the success of its results). Our CNN is capable of handling not only fixed sets of Hubbard model parameters but also site-specific parameters, where each quantum dot may exhibit unique values for coupling constants, intra and inter site Coulomb repulsions, and single particle energies. We explore the prediction of single-parameter per site disorder as well as simultaneous disorder across all parameters. This method marks a significant advance from prior approaches that focused solely on on-site ($\epsilon_i$) disorders \cite{hensgens2017quantum}. Additionally, unlike previous studies focusing only on two dots, we investigate tuning three, four, and five quantum dots concurrently, thereby accounting for crosstalk effects typically overlooked when only two quantum dots are tuned at a time. To our knowledge, this is the first successful attempt to determine site-specific disordered extended Hubbard model parameters for more than two quantum dots including all the coupling constants and intra/inter-site Coulomb repulsions. It is also the first to address site-specific disordered gate voltages/on-site potential for more than three quantum dots. Our CNN enables the possibility for effectively fully automated tuning of quantum dot based systems removing the time consuming manual tuning process currently performed every time the parameters drift due to disorder noise.

Beyond this, we also introduce a "no trust" method for verifying the output of our neural network, which offers a significant advantage over other neural network approaches by preventing the possibility of hallucination and enabling conclusive experimental empirical  confirmation of the model's predictions. Finally, we conclude by generalizing our approach, discussing how our method can be applied to solve disorder in a broad range of physics problems while utilizing the "no trust" confirmation scheme we have {proposed} in this work and previously. The paper is organized as follows:
\begin{enumerate}
\renewcommand{\labelenumi}{\Roman{enumi}.}
\renewcommand{\labelenumii}{\Alph{enumii}.}
\item Methods
    \begin{enumerate}
        \item Model: Describes the physical model, its Hamiltonian, disorder implementation, and simulation methods.
        \item Current Tuning Process:  {Provides background} briefly describ{ing} the {currently used} process  for tuning quantum dots using charge stability diagrams.
        \item Training Data: Explains training data generation and setting up the disorder for the machine learning problem.
        \item Machine Learning: Discusses the deep learning CNN architecture and training process.
    \end{enumerate}
\item Results
    \begin{enumerate}
        \item Single Parameter Learning: Demonstrates the neural network's ability to accurately identify disorder in each Hubbard parameter individually with high precision.
        \item Many Parameter Learning: Demonstrates the ability to accurately identify Hubbard deviations when disorder affects all parameters.
        \item No Trust Verification { Scheme}: Details a method to verify machine learning results without relying on the neural network output, addressing the issue of hallucination. 
        \item Costs: discusses the computational costs of our method.  
        \item Generalization: Presents a framework generalization that allows its use for solving broader disorder problems in physics.
    \end{enumerate}
\item Conclusion
\end{enumerate}



    










\section{Methods}
\subsection{Model}
We consider a ring of quantum dots within semiconductor platforms such as silicon or GaAs heterostructures, which have been accurately modeled by the generic Hubbard model \cite{barthelemy2013quantum,byrnes2008quantum,hensgens2017quantum}. The generic extended Hubbard model is expressed as follows:
\begin{multline}
H=-\sum_{<i,j>, \sigma}t_{ij} \left(c^\dagger_{i\sigma}c_{j\sigma}+h.c.\right)-\sum_i \epsilon_i n_i\\ +\sum_{<i,j>}V_{ij}n_in_j +\sum_i \frac{U_i}{2} n_i (n_i -1) 
\end{multline}

Here, $n_i=n_{i\uparrow}+n_{i\downarrow}$ is the number operator, and $c_{i\sigma}$ is the fermion annihilation operator, with $i$ enumerating the site locations and $\sigma$ the spin degree of freedom (up/down spins). The parameter $t_{ij}$ represents the hopping amplitude between sites, $\epsilon_i$ is the single-particle energy, $V_{ij}$ denotes the inter-site Coulomb repulsion (where i.j indicate different sites), and $U_i$ stands for the on-site repulsion (essentially the diagonal $V_{ii}$ component of $V_{ij}$). We treat the disorder in our model as an additive unknown random deviation from some expected nominal value in these Hubbard model parameters. The parameters with disorder-induced deviations are labeled as $\tilde{t_{ij}}=t_{ij}+\delta t_{ij}$, $\tilde{V_{ij}}=V_{ij}+\delta V_{ij}$, $\tilde{U_i}=U_i+\delta U_i$, and $\tilde{\epsilon_i}=\epsilon_i+\delta \epsilon_i$. For each parameter $\chi$, $\tilde{\chi}$ is the new (disordered) Hubbard model parameter incorporating the disorder deviation $\delta \chi$. (This separation only implies that the desired configuration is without the disorder deviations whereas the quantities with deviations are what the sample has to deal with and the tuning process must bring the system back to the undeviated parameters.) These deviations can be considered specific detunings for each parameter. Only nearest neighbor interactions (on a ring) will be considered such that $\tilde{V}_{ij}=\tilde{t}_{ij}=0$ for $j\neq i+1$. This is a non-essential approximation and is made only because it applies very well to semiconductor quantum dot qubit platforms-- it is straightforward to relax this approximation if necessary. Similarly, the ring configuration with periodic boundary conditions is not a limitation or an approximation, the corresponding linear chain with periodic boundary conditions should manifest very similar behavior and is numerically easier to handle.

Incorporating the disorder into our model leads to the disordered quantum dot Hamiltonian as follows:

\begin{multline}
H=- \sum_{\langle i,j \rangle, \sigma}\tilde{t}_{ij} \left(c^\dagger_{i\sigma}c_{j\sigma}+h.c.\right)-\sum_i \tilde{\epsilon}_i n_i\\ +\sum_{<i,j>}\tilde{V}_{ij}n_in_j +\sum_i \frac{\tilde{U}_i}{2} n_i (n_i -1) 
\end{multline}

The objective is to ascertain the deviations in each Hubbard model parameter, which effectively determines the parameters of the disordered Hamiltonian. The system is considered at a finite (low) temperature $\frac{k_B}{\beta}=0.005 \langle U\rangle$ (with $\langle U\rangle$ being the average U value over the sites), with site-dependent chemical potentials $\mu_i$ being variable. Our measurements are based on expectation values $\langle n_i\rangle=\text{Tr}(\rho n_i)$, where $\rho=e^{-\beta (H -{\sum_i}\mu_i n_i)}$, and are taken across an array of different $\mu_i$ values. Experimentally, these measurements are facilitated by applying a site-dependent gate voltage. All sites are measured simultaneously, resulting in an occupation vector $\vec{n}=(\langle n_1\rangle, \cdots , \langle n_N\rangle)$. Numerically, the simulation was carried out through exact diagonalization using the Quspin package \cite{weinberg2017quspin}. To account for thermal effects, the bottom 10 lowest energy states were retained for generating $\rho$, an amount that is more than adequate (for describing the typical experimental situation) given the low experimental temperature setting. Note that as long as the temperature is low, its precise value is irrelevant. { Even in the case of larger systems with small coupling constant, a lower T say $\frac{k_B}{\beta}=0.0026 \langle U\rangle$, which is experimentally achievable \cite{hensgens2017quantum}, should be sufficient to maintain an effectively pure state for significantly larger system sizes than we consider.)}

\subsection{{ Current} Tuning Process}
\begin{figure}[h]
     \centering
     \begin{subfigure}[b]{0.9\linewidth}
         \centering
         \includegraphics[width=\textwidth]{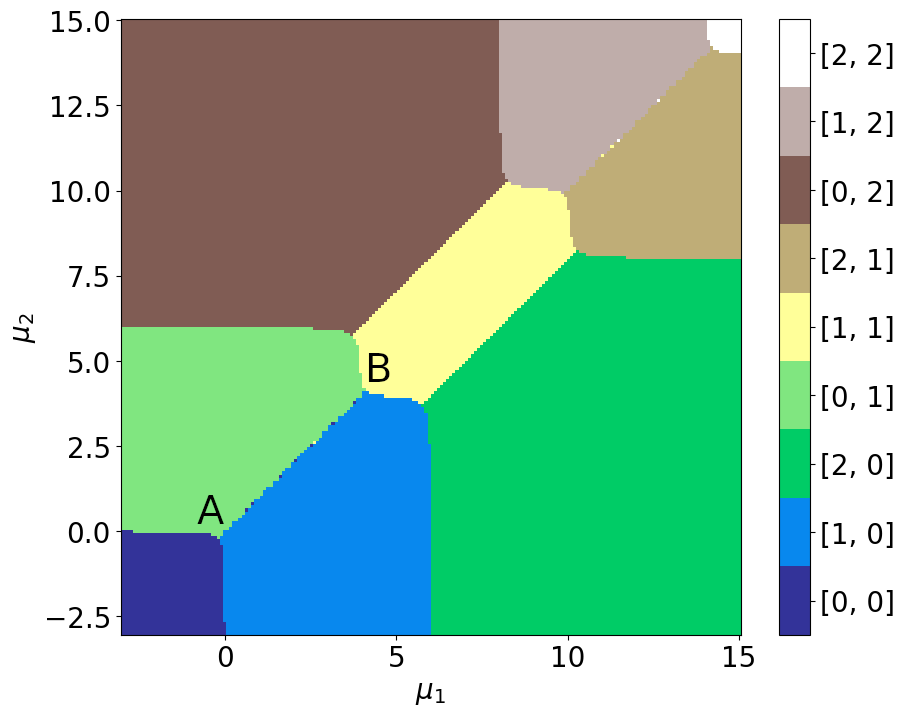}
         \caption{}
     \end{subfigure}
     \caption{{ Standard charge stability diagram showing the most probable states calculated at $U_1=U_2=6$, $t=0.1$, $V_{12}=2$ and $\epsilon_i=0$}  }
     \label{fig:StandardDiagram}
    \end{figure}

The current{ly used} method for determining the Hubbard model parameters for quantum dots involves assessing the parameters for each pair of dots by examining their charge stability diagrams. This approach,  { unlike ours, }requires assuming that the effects of other dots can be either removed or rendered negligible. Charge stability diagrams depict the stable electronic configurations (or expected occupation numbers) across different sites as a function of the chemical potentials (or experimentally, gate voltages). In the scenario of two quantum dots, the Hubbard parameters' influence on the stability diagrams behaves in a predictable manner, making it possible to deduce these parameters. For example, the parameter t visually curves the boundaries within the charge stability diagram. The value of t between two quantum dots can be inferred from the boundary between the (0,0) occupation phase and the (0,1) or (1,0) occupation phases. {Experimentally, these boundaries are peaks in the charge addition spectrum.} The boundary is represented by the equation $\mu_1*\mu_2 = t^2$, leading to the triple point labeled A, where $\mu_1(A)=\mu_2(A)=-t$ \cite{yang2011generic}. Assuming t is sufficiently small (which is the generic situation), $U_i$ can initially be approximated within the stability diagram, based on the energy needed to add a single electron to the i-th dot. If it is known that $U_1=U_2=U$, then $V_{12}$ for two quantum dots can be determined from the position of a triple point within the charge stability diagram \cite{wang2011quantum}. Triple points are intersections of three distinct phases. The necessary triple point, labeled B, involves [(1,0),(0,1),(1,1)] occupations and occurs at $\mu_1(B)=\mu_2(B)=t+\frac{(U+V_{12})}{2}-\sqrt{4t^2+\frac{(U-V_{12})^2}{4}}$. {( See Fig. \ref{fig:StandardDiagram})} In situations where $U_1 \neq U_2$, or when t is not small enough to simplify the $U_i$ determination, numerical fitting should be done manually to determine $V_{12}$, highlighting a primary limitation in current methodologies \cite{sarma2011hubbard}. For more details on this background, we refer the reader to Refs. \cite{hensgens2017quantum,sarma2011hubbard,wang2011quantum}. 

The site-dependent disorder, $\tilde{\epsilon}_i$, has been investigated for tuning three quantum dots in a previous work \cite{hensgens2017quantum}, where a method of least squares with phase diagram transition lines (identified using an edge-finding algorithm) is employed to fit the parameters. This method proves effective but is limited to small $t/U<0.15$, as it fails at larger coupling values because the phase transition lines cease to be sufficiently straight { and $n_i$ cease to be good quantum numbers making the boundaries diffuse.} In summary, the current method for determining the Hubbard model parameters relies heavily on manual fitting beyond a certain point, using two or at most three quantum dots, which is notably inefficient.

\subsection{Training Data}

The training data for the machine learning model consists of performing occupation expectation $\vec{n}$ measurements across various measurement configurations. Each configuration specifies a list of $\mu_i$ for every site. However, for practical purposes, within each nearest neighbor pair of quantum dots $i$ and $i+1$, we assign the parameters as follows: {$\mu_i$}, $\mu_{i+1}=-\mu_i$, and for any $\mu_{j\neq i}=\mu$. We then independently vary $\mu$ and $\mu_i$ for all sites in our model and across every pair of nearest neighbor sites $i$ and $i+1$, using periodic boundary conditions. (Using open boundary conditions makes the machine learning procedure somewhat easier numerically, with similar results-- the periodic/open boundary conditions physically correspond to 2D/1D qubit configurations, { namely 4 dots with periodic boundaries is equivalent to 2x2 dots with an open boundary.})  Specifically, $\mu$ is adjusted between $[-0.5W,2.5W]$ with 50 steps, and $\mu_i$ is varied between $[-0.6\langle U\rangle,+0.6\langle U\rangle]$ with 20 steps, where $W=\frac{1}{L}\left(\sum_i \langle U_i \rangle +\langle V_{i,i+1}\rangle\right)$ acts as the normalization factor. These normalization factors are based on their deviation-free values, consistent across all disorder realizations, although this adjustment is merely a linear factor and not essential. { It should be noted that this measurement space is significantly less than the available observation space, this being sufficient is surprising and determined in this work posteriori.} The various measurement configurations are recorded in a matrix K, where each row records a specific setup of experimentally tunable parameters, labeled as j. The structure of the K matrix is presented below:

\begin{equation}
K^j=[\mu^j,\mu_i^j,i^j_{site}]
\end{equation}

{ Where $i^j_{site}$ records the site of $\mu_i^j$.} For each measurement configuration (or row $K^j$), the expected occupation number of each quantum dot is used to form an input matrix $X$. This $X$ matrix, formed from experimental measurements, will be utilized by the neural network to determine the Hubbard parameters{, though i}n our case, of course, these 'experimental measurements' producing the data are generated by the actual simulations of the disordered Hubbard model{:}

\begin{equation}
X^j=[n_1^j, n_2^j, n_3^j, \cdots]
\end{equation}

The output vector of the machine learning scheme, which in this case also serves as the input vector to the quantum simulation, will be denoted as the vector $Y$. The vector $Y$ comprises a list of all the disorder deviations in the Hubbard model parameters. This can be written as follows:

\begin{equation}
Y=\begin{bmatrix} \vec{\delta \epsilon} \\ \vec{\delta V} \\ \vec{\delta t} \\ \vec{\delta U} \end{bmatrix}
\end{equation}

Through exact diagonalization, it is possible to accurately simulate, for a limited number of dots (perhaps up to 10-20 in principle limited by the computational resources underlying the exact diagonalization ), the occupation expectations matrix $X$ for the different measurement configurations included in our $K$ matrix and a given set of Hubbard model parameters $Y$. We can express this as a function as follows:

\begin{equation}
f_{Gen}(Y;K)=X
\end{equation}

This function serves as our generator, responsible for creating our training data. To produce a large volume of data for neural network training, we input numerous randomly generated realizations of $Y$ and generate their respective $X$. It is important to note that all different realizations of $Y$ share the same $K$ matrix. Therefore, our goal is to find a function capable of taking $X$ and $K$, and returning $Y$. In other words, we aim to invert our generator function to achieve the following:

\begin{equation}
f^{-1}_{ML}(X;K)=Y
\end{equation}

\begin{figure}[h]
     \centering
     \begin{subfigure}[b]{1.0\linewidth}
         \centering
         \includegraphics[width=\textwidth]{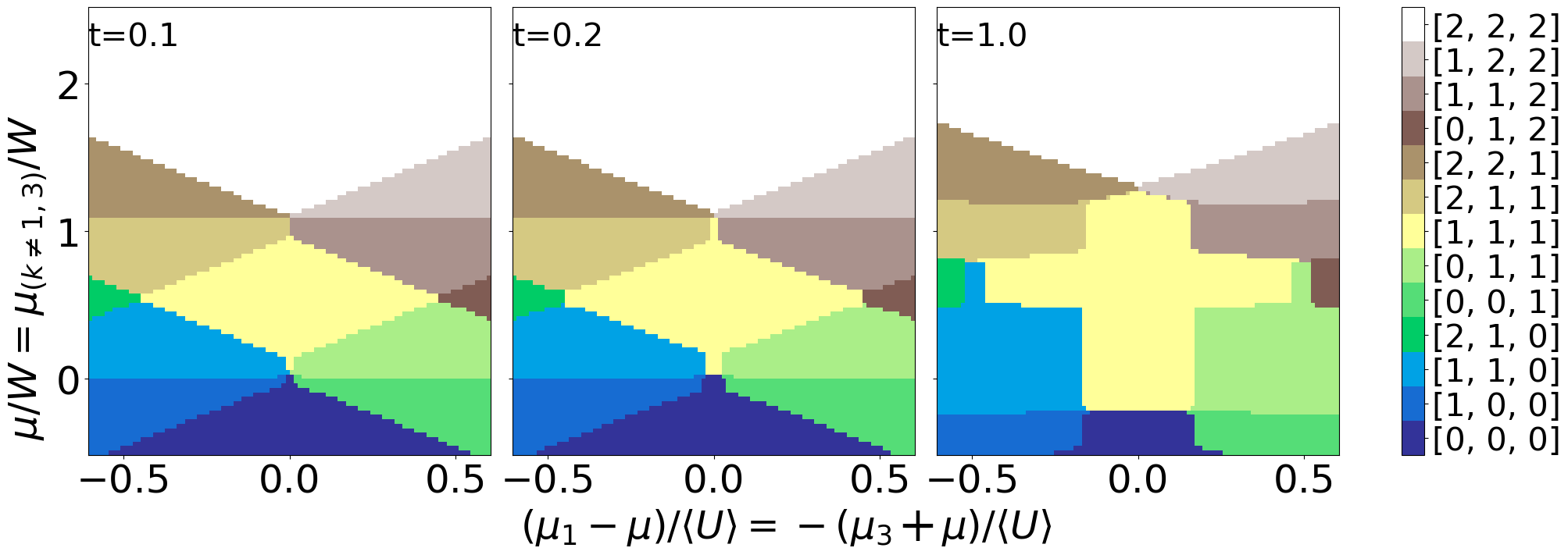}
         \caption{}
     \end{subfigure}
        \begin{subfigure}[b]{1.0\linewidth}
         \centering
         \includegraphics[width=\textwidth]{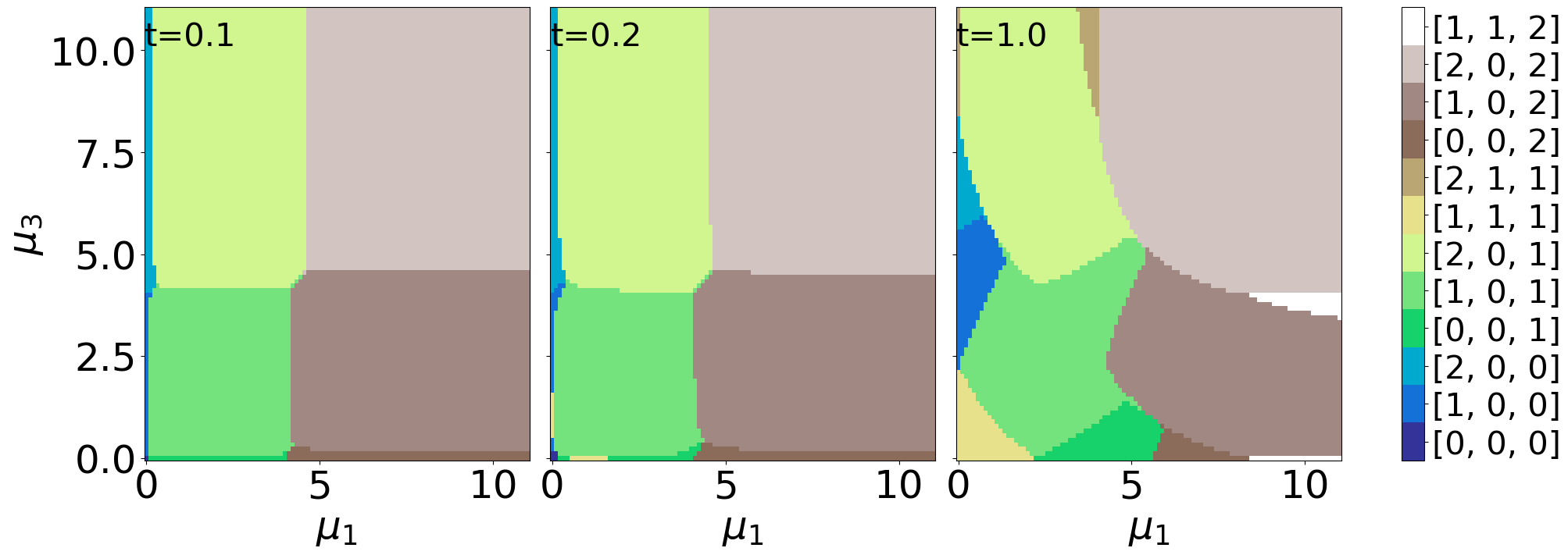}
         \caption{}
     \end{subfigure}
     \caption{(a) Disorder free example of training data. The chemical potentials $\mu_1=-\mu_3$ (rescaled by $\langle U\rangle$ and shifted by $\mu$) vs. $\mu_{k\neq 1,3}=\mu$ (rescaled by $W=\frac{1}{L}\left(\sum_i \langle U_i \rangle +\langle V_{i,i+1}\rangle\right)$)  (b) Disorder free example of standard charge stability diagram. Both (a-b) are for the 3 quantum dot pristine system with t=0.1, 0.5, 1.0. The Hubbard Parameters were fixed $U=4$, $V_{i,i+1}=0.2$, $\epsilon_i=0$. The color bar represents the most probable state at the chemical potential configuration labelled by the plot.}
     \label{fig:PristineTraining}
    \end{figure}

Putting everything together for clarity, our input X is a series of a version of charge stability diagrams. The neural network takes in a series of nearest neighbor charge stability diagrams to determine the Hubbard model parameters for a ring of quantum dots. { This input is essentially the same as what is already being used to experimentally tune dots without any additional processing \cite{hensgens2017quantum}.}  { A pristine example of the diagram can be seen in Fig. \ref{fig:PristineTraining}.}

\subsection{Machine Learning}

\begin{figure}[h]
     \centering
     \begin{subfigure}[b]{1.0\linewidth}
         \centering
         \includegraphics[width=\textwidth]{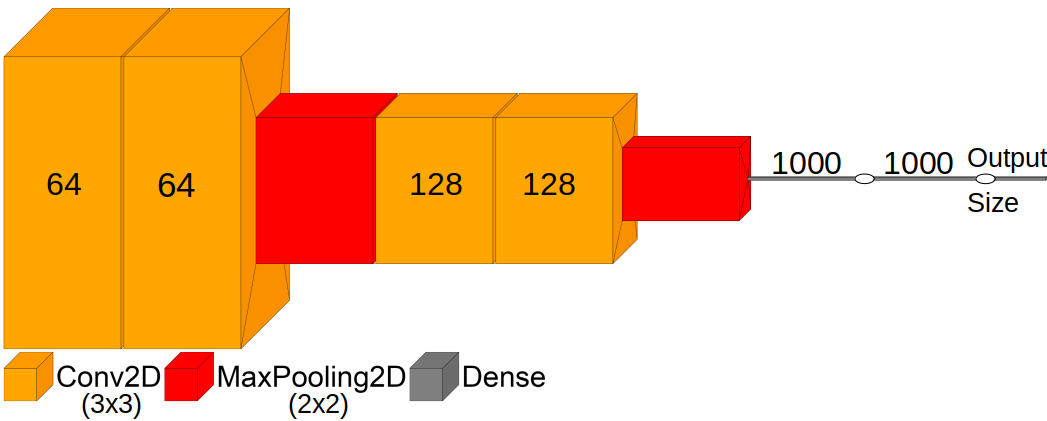}
         \caption{}
     \end{subfigure}
     \begin{subfigure}[b]{1.0\linewidth}
         \centering
         \includegraphics[width=\textwidth]{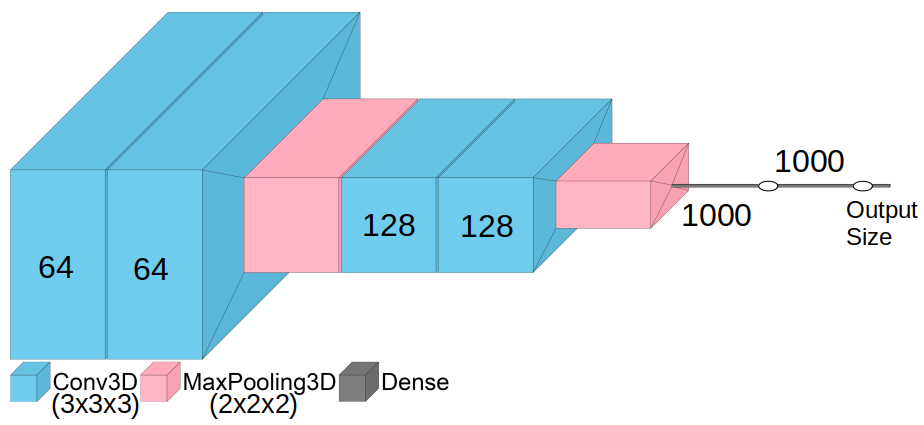}
         \caption{}
     \end{subfigure}
     \caption{Diagrams of the two different neural networks used. Both machine learning models consist of a convolutional neural network (CNN) portion followed by two dense layers. The CNN is composed of two sets of convolutional layers: the first set has two layers with 64 filters, followed by a second set with two layers of 128 filters. Each set of convolutional layers, consisting of either (a) two 2D convolutional layers with a 3x3 kernel or (b) two 3D convolutional layers with a 3x3x3 kernel, is followed by either (a) a 2D max pooling layer of size 2x2 or (b) a 3D max pooling layer of size 2x2x2. The networks conclude with two dense layers of size 1000 and one dense layer equal to the size of the output. In all cases, the network employs a RELU function to introduce non-linearity.}
     \label{fig:NNDiagram}
    \end{figure}

The neural network we use utilizes a particularly simple convolutional neural network (CNN), based on a greatly simplified version of AlexNet \cite{krizhevsky2012imagenet}. CNNs have proven especially effective in solving problems where the input is akin to a vision problem, with some locality in the data. In our case, the input is a series of charge stability diagrams, which are presently individually visually inspected to determine parameters, making the application of a CNN to this problem quite natural since a CNN is an extremely efficient technique for implementing 'vision'.

Two different but very similar architectures {are} employed in the current work: one with 2D convolutional layers and another with 3D convolutional layers. These setups, depicted in Fig. \ref{fig:NNDiagram}, will be referred to as the 3D type and the 2D type, respectively. In both types, the neural network consists of two sets of convolutional layers, each containing two layers. In the 2D (3D) type, the first set includes two 2D (3D) convolutional layers with 64 filters and a kernel size of 3x3 (3x3x3), while the second set is identical but with 128 filters. A 2D (3D) max pooling layer with a size of 2x2 (2x2x2) is placed between each set. Following the convolutional sets in both types are two dense layers with 1000 nodes each and a final dense layer equal to the output size. This relatively simple neural network configuration is notably efficient in predicting the Hubbard parameters. For the 2D type, the input {is} in the form of charge stability diagrams, as previously mentioned, but these were reshaped into a 2D grid based on the $\mu$ and $\mu_i^j$ values, where the channels represent the expected occupation values of different quantum dots. However, this setup does not include the site number $i_{site}$ in the input. To incorporate this parameter, the 2D grid for each different site pair {is} appended sequentially to the 2D grid, resulting in input data that took the form of a grid of occupation numbers based on the $\mu$ and $\mu_i^j$ values, where $i=1, 2 \cdots$, all appended together along the $\mu^j$ axis. This peculiar setup was chosen for the 2D type to facilitate the use of a 2D CNN and to make the input more symmetrical. In the 3D type, this appending {is} unnecessary as the convolution spanned all three axes: $\mu$, $\mu_i^j$, and $i_{site}$. The 2D type architecture {is} utilized for predictions involving deviations in a single parameter, whereas the 3D type architecture { is} applied to predict deviations in multiple parameters. For both types, 90\% of the generated disorder realizations were used for training, and the remaining 10\% were reserved for testing and validation. The validity of our model {is} assessed by calculating the $R^2$ (coefficient of determination) and the root mean squared (RMS) error for our predictions { of the deviations} on the validation data. The training (90\%) and testing (10\%) data were selected randomly with no bias.

\section{Results}
\subsection{Single Parameter Learning}

To start, let us examine the simpler scenario where there is an error in only one Hubbard model parameter. Specifically, we will look at the case of three quantum dots with an error or deviation only in $\delta \epsilon$. The scenarios involving one and two quantum dots were not considered, as these situations are well understood, and the use of machine learning would be trivial and an overkill.  For the scenario where only $\delta \epsilon_i \neq 0$, we assume $\delta V_{i,i+1}=\delta t_{i,i+1}=\delta U_{i}=0$. In this model, $\delta \epsilon_i \in [-0.5,0.5]$ is chosen from either a Gaussian or a normal distribution with an equal probabilities {to increase the complexity in the data and demonstrate our methods robustness). When applicable all deviations will use both types of distributions except $\delta t_{i,i+1}$ which will only use uniform.} The remaining parameters are kept constant at $V_{i,i+1}=0.2$, $t_{i,i+1}=1$, and $U_i=4$. (Unless otherwise stated specifically, we use the nearest neighbor hopping t as the energy unit.)

To further { illustrate} the validity of our method, we test whether our predictions for the Hubbard deviations can be used to regenerate the stability diagrams. This is crucial because it is not immediately clear if the collection of charge stability diagrams we feed in can uniquely determine the Hubbard parameters in the first place since there is no 'inverse scattering' theorem proven for such a scenario. Nonetheless, a match of the charge stability diagrams would { illustrate} how well our predictions capture the physics. Solving for 2 or 3 quantum dots is how this issue is currently addressed experimentally, requiring manual effort and certain restrictions on the parameters. Considerable time is spent periodically to re-tune the dots by determining and then correcting for $\delta \epsilon_i$. With our machine learning model, we find we can solve this problem efficiently with high accuracy. Our model, in the case of three quantum dots with only $\delta \epsilon_i$ errors, achieves an $R^2=0.99996$ with an RMS error of $R_{MS}(\delta \epsilon)=0.00183$ at each site. { Both $R^2$ and $R_{MS}$ metrics presented are in terms of the Hubbard parameter deviations to assess the validity of our method.} We are not just finding $\delta \epsilon_i$ for individual sites but for the entire position-based function. The results, shown in Fig. \ref{fig:3SingleParamEpsilonMu2}, illustrate that deviations in gate voltages primarily result in shifts in the phase boundaries, making $\delta \epsilon$ seemingly the easiest parameter to determine. This suggests that, at least for site-specific $\epsilon$ errors within our parameter range, the convolutional neural network can uniquely identify the $\delta \epsilon$ disorder by using merely a series of nearest neighbor charge stability diagrams. This outcome is significant {\comment{(and utterly non-obvious)}} as it could enable the automation of re-tuning this single parameter, removing the need for manual intervention, which is not significantly restricted by the smallness of $t$ as mentioned in \cite{hensgens2017quantum}.

However, solving $\delta \epsilon_i$ for 3 quantum dots beyond the t range, while beneficial for automation and possibly the speed of determination, does not address a problem that is not already regularly solved manually in the experimental setting. The real machine learning advantage lies in its ability to solve for other unknown parameters and the fact that it can do so with larger numbers of quantum dots at once. Currently, quantum dots are experimentally tuned two or three at a time, with this aggregate tuning deemed sufficient for the whole device \cite{hensgens2017quantum}. Our method, however, allows many more quantum dots to be tuned at once, providing more accurate and much faster automated tuning, and enabling more parameters to be adjusted. In the case of four quantum dots with errors only in $\delta \epsilon_i$ using the same parameter setup { (see Fig. \ref{fig:4SingleParamEpsilonMu2})}, we achieved an $R^2=0.99995$ and an average prediction error of $R_{MS}(\delta \epsilon)=0.0029$.  This is an impressive result that, unlike previous work, allows four or more quantum dots to be tuned simultaneously. A latter section will demonstrate that this approach is effective for five quantum dots as well, with minimal impact on fidelity. Yet, the capability to tune more dots simultaneously is not the only significant advantage of our method; it also enables tuning of parameters beyond $\delta \epsilon$ that are more challenging (and typically not done manually in the laboratory).

Next, we examine the case of determining site-dependent values of $\delta V_{ij}$, $\delta t_{ij}$, and $\delta U_{i}$, where we assume errors in only one parameter at a time and presume accurate estimations can be achieved for the other parameters. This approach is taken to determine whether our CNN scheme can predict all these parameters or just $\epsilon_i$ errors. In a single parameter error scenario with four quantum dots and approximately 40,000 training realizations, we find that $\delta V_{ij}$ can be predicted very accurately with an $R^2=0.99995$ and $R_{MS}(\delta V_{ij})=0.0009$ and is shown in Fig. \ref{fig:4SingleParamVij}. $\tilde{V}_{ij}$ is assumed to be within a range of $[0,0.4]$. This demonstrates the CNN's ability to accurately predict this parameter for each bond uniquely. In the case of single parameter error $\delta t_{ij}$ at each bond, the CNN predicts the deviation with an $R^2=0.99995$ and $R_{MS}(\delta t_{ij})=0.0243$, and is shown in Fig. \ref{fig:4SingleParamdt}. $\tilde{t}_{ij}$ is assumed to be within a range of $[0.1,10]$. For single error $\delta U_{i}$, the CNN predicts the deviation with an $R^2=0.9988$ and $R_{MS}(\delta U_{i})=0.101$, and is shown in Fig. \ref{fig:4SingleParamUi}. $\tilde{U}_{i}$ is assumed to be approximately within $[0,8]$, ensuring that the correct sign is maintained. In all cases, and thus for all our Hubbard parameters, the neural network accurately predicts the deviation errors from just a series of charge stability diagrams. This is an important {\comment{(and rather surprising as well as useful)}} finding since charge stability diagrams are routinely measured experimentally in quantum dot qubit devices.

\clearpage
\onecolumngrid
\begin{minipage}{\linewidth} 
\begin{figure}[H]
     \centering
     \begin{subfigure}[b]{0.9\linewidth}
         \centering
         \includegraphics[width=\textwidth]{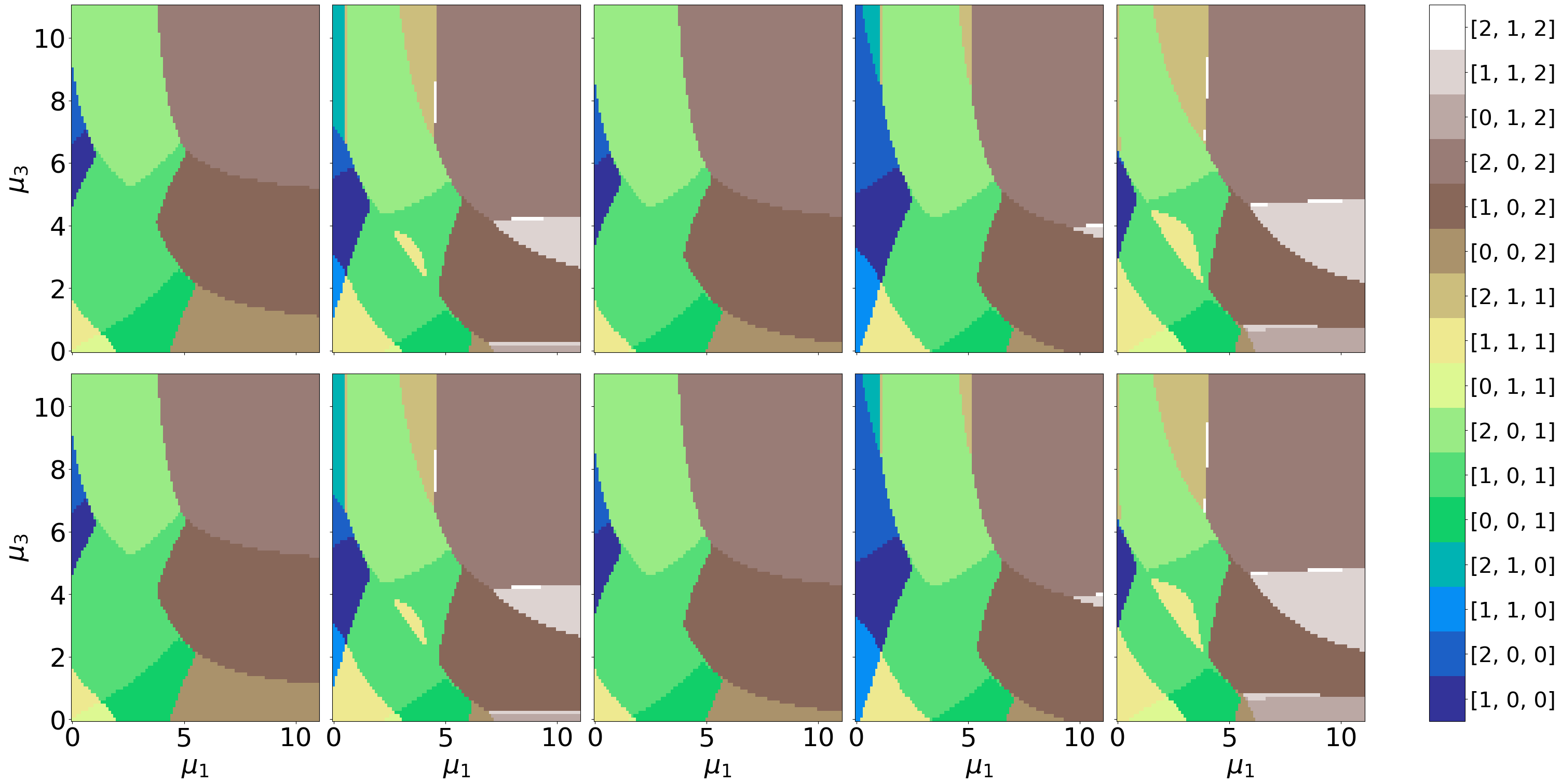}
         \caption{}
     \end{subfigure}
     \begin{subfigure}[b]{0.9\linewidth}
         \centering
         \includegraphics[width=\textwidth]{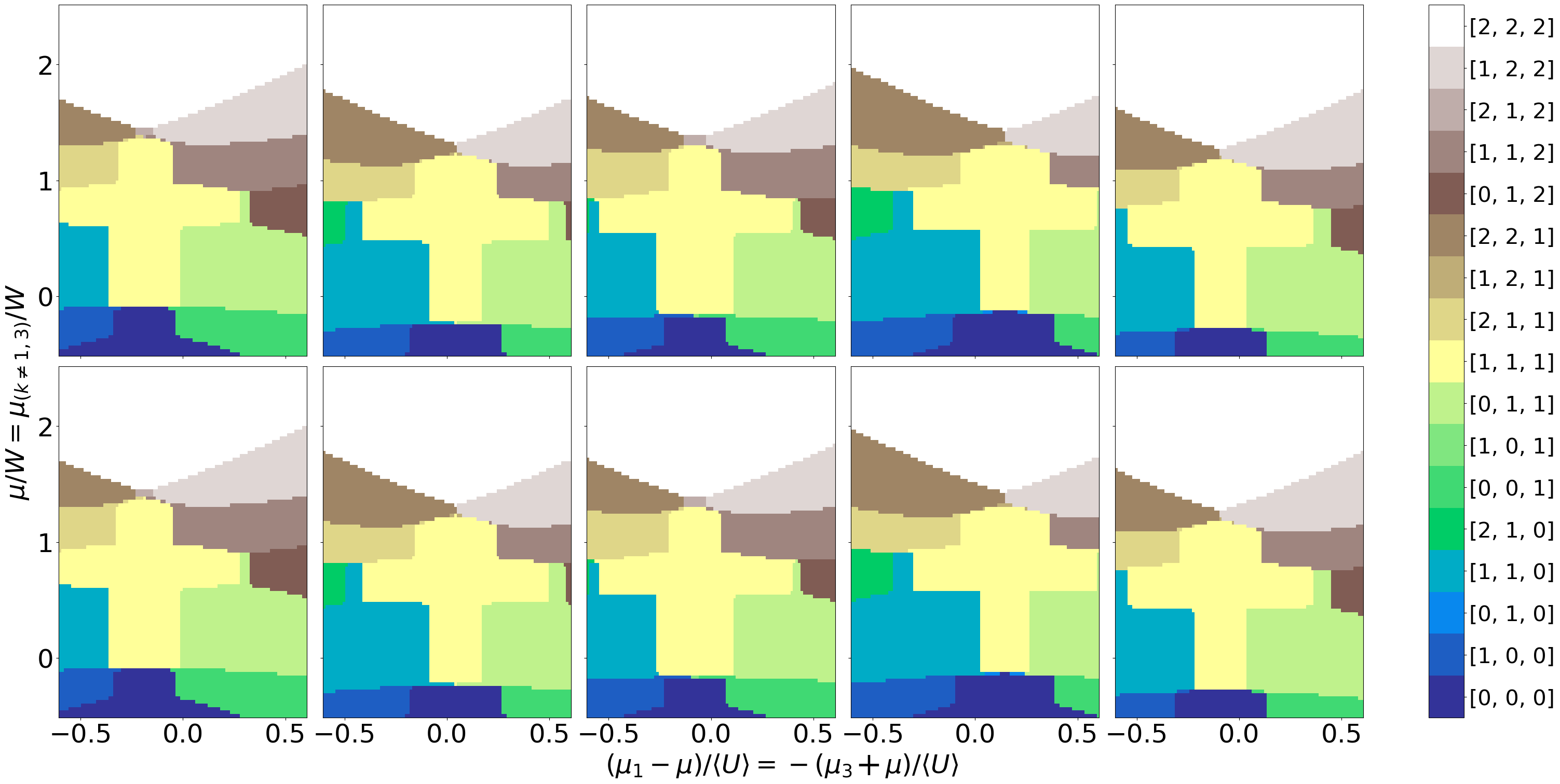}
         \caption{}
     \end{subfigure}
    \caption{3 quantum dot charge stability diagrams for input and expected measurement outcomes from Hubbard model parameters for the prediction of only $\delta{\epsilon_i}$. The root mean squared error in the Hubbard model $\epsilon_i$ parameter was $R_{MS}({\delta \epsilon})=0.00183$ with an $R^2=0.99998$. In both plots the first row is the input charge stability diagrams, namely the most probable state for a chemical potential configuration, and the second row is the expected charge stability diagram given the prediction of the Hubbard model parameters. The error-free model parameters were set to $U=4$, $t=1$ and $V_{i,i+1}=0.2$. The 2D type CNN was used for this model. The columns of these two subplots correspond to the same {randomly selected} representative {validation} samples. (a) Most probable state for input Hubbard parameters (1st row) and predicted Hubbard parameters (2nd row) where $\mu_1$ and $\mu_3$ are independently varied. (b) Most probable state for input Hubbard parameters (1st row) and predicted Hubbard parameters (2nd row) where the chemical potential at each site is $\vec{\mu}=[\mu_1,\cdots,\mu_n]$ with our plot having axis $\mu_1=-\mu_3$ (rescaled by $\langle U \rangle$ and shifted by $\mu$) vs. $\mu_{k\neq 1,3}=\mu$ (rescaled by $W=\frac{1}{L}\left(\sum_i \langle U_i \rangle +\langle V_{i,i+1}\rangle\right)$ where $\langle U\rangle=4$ and $\langle V_{i,i+1}\rangle=0.2$ are the disorder free values respectively), this is similar to the stability diagrams fed into the neural network.}
    \label{fig:3SingleParamEpsilonMu2}
\end{figure}
\end{minipage}

\begin{minipage}{\linewidth} 
\begin{figure}[H]
     \centering
     \begin{subfigure}[b]{0.65\linewidth}
         \centering
         \includegraphics[width=\textwidth]{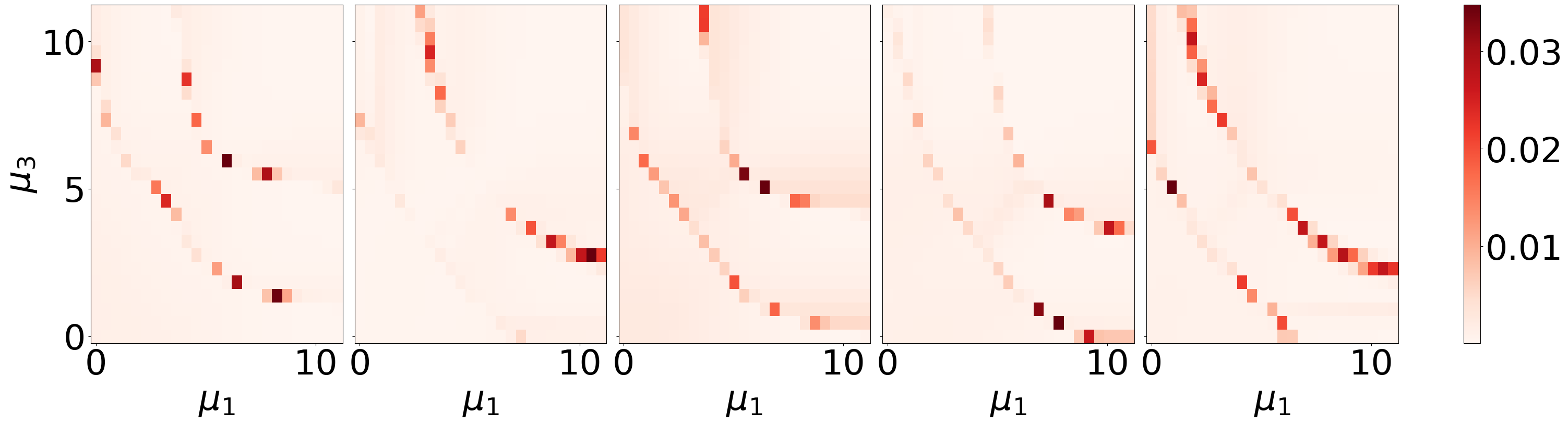}
         \caption{}
     \end{subfigure}
     \begin{subfigure}[b]{0.65\linewidth}
         \centering
         \includegraphics[width=\textwidth]{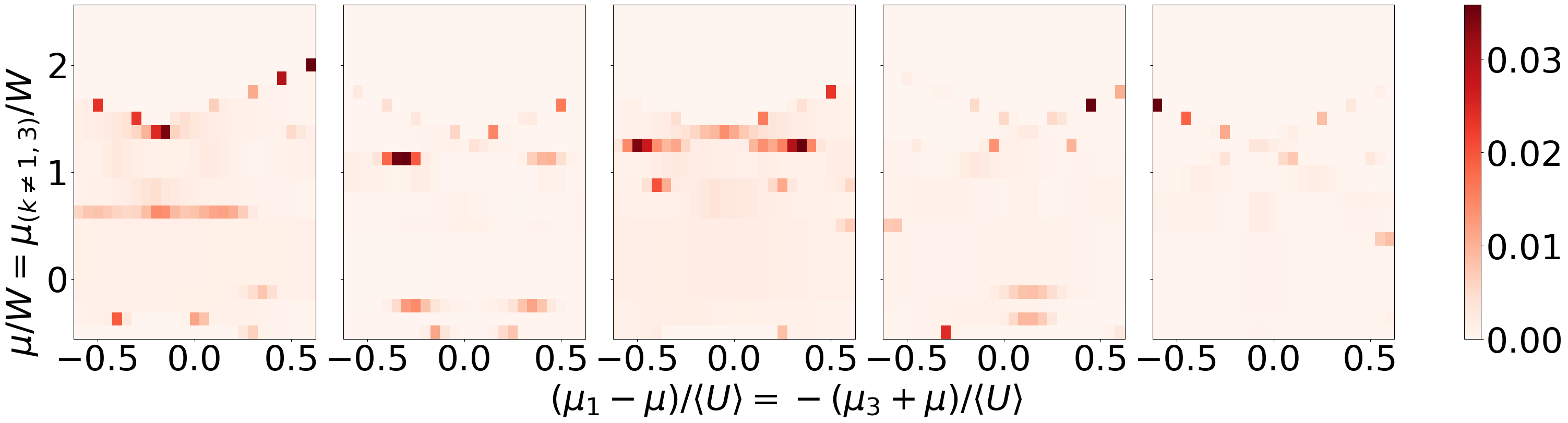}
         \caption{}
     \end{subfigure}

    \caption{Differences in charge stability diagrams for 3 quantum dots, comparing input with expected outcomes from Hubbard model parameters for the prediction of only $\delta {\epsilon_i}$. In particular $n_{error}=||\langle\vec{n}_{input}\rangle-\langle\vec{n}_{predicted}\rangle||$ is plotted in both plots where $\langle\vec{n}_{input}\rangle$ is the occupation expectation vector for all sites of the input data and $\langle\vec{n}_{predicted}\rangle$ is the expected occupation expectation vector given the prediction of the Hubbard model parameters. The 2D type CNN was used for this model. The columns of both plots correspond to the charge stability diagrams in Fig. \ref{fig:3SingleParamEpsilonMu2}. The error-free model parameters were set to $U=4$, $t=1$ and $V_{i,i+1}=0.2$. (a) $n_{error}$ between input Hubbard parameters and predicted Hubbard parameters where $\mu_1$ and $\mu_3$ are independently varied. (b) $n_{error}$ between input Hubbard parameters and predicted Hubbard parameters where the chemical potential at each site is $\vec{\mu}=[\mu_1,\cdots,\mu_n]$ with our plot having axis $\mu_1=-\mu_3$ (rescaled by $\langle U \rangle$ and shifted by $\mu$) vs. $\mu_{k\neq 1,3}=\mu$ (rescaled by $W=\frac{1}{L}\left(\sum_i \langle U_i \rangle +\langle V_{i,i+1}\rangle\right)$ where $\langle U_i\rangle=4$ and $\langle V_{i,i+1}\rangle=0.2$ are the disorder free values respectively).}
    \label{fig:E3SingleParamEpsilonMu}
\end{figure}
\end{minipage}  
\newpage
\twocolumngrid

\subsection{Many Parameter Learning}
{\comment{ In the last section, it was shown that high-fidelity predictions of individual Hubbard parameters can be achieved using just a few stability diagrams, assuming that deviations for all but one Hubbard parameter are small. This approach assumes accurate predictions of the deviations in the remaining parameters. However,}}{Building on high-fidelity predictions of the previous section,}  in practical experiments, all parameters typically exhibit errors that need to be addressed simultaneously. To meet this challenge, we introduce deviations from the expected values for all our Hubbard parameters. Our machine learning method, leveraging nearest neighbor stability diagrams, is capable of determining all the Hubbard parameters at once, without any initial knowledge of any specific parameter { beyond a very rough range/scale necessary for generating training data\comment{(or no knowledge except a reasonable range)}}. In our investigations involving multiple parameters, we train our neural network with 10,000 to 20,000 training realizations. This is a reduction from the 40,000 realizations used in single-parameter scenarios, motivated by our findings that so many realizations are not necessary for achieving high fidelity and by the increased numerical demand of the simulations when all parameters exhibit disorder. Nonetheless, it is apparent that including more training realizations could further enhance the (already pretty high) fidelity.

To begin, consider the 3 quantum dot case, with a more limited parameter range of $\tilde{\epsilon}_i \in [-0.5,0.5]$, $\tilde{V}_{ij} \in [0,0.4]$, $\tilde{t}_{ij} \in [0.01,0.25]$, and $\tilde{U}_{i} \in [3,5]$. To set a scale for the system, $t_{12}=0.125$ is fixed. We find that the system can predict the Hubbard deviations with an $R^2=0.9987$ and RMS errors of $R_{MS}({\delta \epsilon_{i}})=0.0128$, $R_{MS}(\delta V_{i,j})=0.0190$, $R_{MS}({\delta t_{i,j}})=0.0119$, and $R_{MS}(\delta U_{i})=0.0358$. Although this prediction is not as precise as in the single parameter regime, it remains very accurate. {\comment{This outcome strongly suggests that the CNN, when fed with charge stability diagrams, can predict all the Hubbard model parameters simultaneously.}} The figure for this can be viewed in Fig. \ref{fig:3AllParamFullS}. An assessment of the actual error within the stability diagrams due to inaccuracies in the predictions was also performed. This assessment is crucial because our plots only display the most probable state, and it is possible that the differences between the predicted and input stability diagrams are not as significant as they appear, potentially being an artifact of this selection process. To explore this, we plot the magnitude of the error in the occupation expectation vectors between the input and predicted Hubbard model parameters as $||\langle\vec{n}_{input}\rangle-\langle\vec{n}_{predicted}\rangle||$ in { Fig. \ref{fig:E3AllParamFullS} and Fig. \ref{fig:E3SingleParamEpsilonMu}}. It was found that the errors are concentrated at the boundaries, which could be due to the low resolution of the stability diagrams inputted into the neural network. This may also result from the majority of information about the Hubbard parameters in the charge stability diagram being concentrated within the phase boundaries. This suggests that our grid scheme of generating the charge stability diagrams is sub-optimal, and utilizing a more sophisticated edge-finding method to extrapolate the diagrams to a higher resolution could lead to significant improvements. The specific impact of increasing the resolution will be discussed later. Nonetheless, these errors are relatively minor, with the stability diagrams being well reproduced from the predicted Hubbard model parameters.

Transitioning to the main many parameter regime, we expand $\tilde{t}_{ij} \in [0.1,2]$ and establish a scale for the system by fixing $t_{12}=1$. It is crucial to recognize that this main many parameter regime is still more constrained than the single-parameter predictions for $\delta U_{i}$ and $\delta t_{ij}$. Despite a decrease in fidelity within this expanded range, the system still achieves an $R^2=0.9977$ with errors of $R_{MS}({\delta \epsilon_{i}})=0.0239$, $R_{MS}(\delta V_{i,j})=0.0164$, $R_{MS}({\delta t_{i,j}})=0.0121$, and $R_{MS}(\delta U_{i})=0.0373$ { (see Fig. \ref{fig:3AllParamFull} and Fig. \ref{fig:E3AllParamFull})}. This shows that increasing our t range, which significantly increases the complexity of the problem, is still able to be handled within some sufficiently small (but still relative to experiment large) range. 

Expanding to a larger number of quantum dots, we first examine the case involving 4 coupled quantum dots with deviations in all parameters. We achieved an $R^2=0.9961$ with errors of $R_{MS}({\delta \epsilon_{i}})=0.0280$, $R_{MS}({\delta V_{i,j}})=0.0205$, $R_{MS}({\delta t_{i,j}})=0.0193$, and $R_{MS}({\delta U_{i}})=0.0512$. This represents only a slight reduction in the fidelity of our predictions, indicating that the tuning approach remains effective for a larger number of quantum dots. The charge stability diagrams are illustrated in Fig. \ref{fig:4AllParamFull}, with the corresponding stability diagram errors depicted in Fig. \ref{fig:E4AllParamFull}.

Our method also performs well in the case of 5 quantum dots with deviations across all parameters. We managed to achieve an $R^2=0.995$ with errors of $R_{MS}({\delta \epsilon_{i}})=0.0321$, $R_{MS}(\delta V_{i,j})=0.0231$, $R_{MS}({\delta t_{i,j}})=0.0254$, and $R_{MS}(\delta U_{i})=0.0574$. Once again, this shows only a minor decrease in prediction fidelity, underscoring that the tuning method is viable for an increasing number of quantum dots. The charge stability diagrams for this scenario are presented in Fig. \ref{fig:5AllParamFull}, with the respective errors shown in Fig. \ref{fig:E5AllParamFull}. These results suggest that the machine learning scheme can automatically tune as many quantum dots as one has computational resources for training data generation, and it can be accomplished with a relatively simple neural network. Given that the Hubbard model is a strongly interacting system, the generation of training data is the primary bottleneck by a wide margin, while machine learning resources, by comparison, do not significantly scale with the number of dots. Ideally, the number of quantum dots one needs to tune simultaneously depends on the effect neighboring quantum dots have on the local state of the system. Bearing this in mind, it is not seemingly necessary when tuning a large system to arbitrarily scale up the number of dots tuned simultaneously { see Section \ref{S:Cost}}. {\comment{In fact, the 5-dot system achieved in our work should describe well all current and near-future quantum dot qubit platforms. When used on actual experiments, the effective Hubbard model parameters will have to be generated first by initially fitting to the experimental data. The machine learning procedure outlined in our work then can be effectively applied to the experiments.}}

Our neural network faces its greatest challenge in accurately predicting the Hubbard model parameters over a very broad $\tilde{t}_{ij}$ range. The curvature induced by $t$ in the charge stability diagrams might complicate the network's task of discerning other parameters, likely necessitating more training data due to the system's increased complexity. When $\tilde{t}$ ranges from $\tilde{t}\in [0.1,10]$, significant errors are found in the predictions of parameters other than $\tilde{t}$. However, this issue remains solvable, as even when predictions for other parameters "fail," the disordered $\tilde{t}$ can still be predicted with high accuracy. For scenarios requiring a large $t$ range, we recommend a sequential procedure, initially generating a prediction for $\tilde{t_i}$ using the full range, then creating training data and training a new neural network within a narrower $t$ range centered around the initial prediction. This approach reduces the dominance of $t$ by focusing the training data within a probable range. Alternatively, generating a significantly larger volume of training data from the outset is an option, though it may be less efficient. Additionally, employing a standard scaling operation, such as those available in the scikit-learn package \cite{pedregosa2011scikit}, proves useful when dealing with output parameters of substantially different scales. This technique ensures that the neural network does not prioritize only predicting $\tilde{t}$ by mapping all parameters in an invertible manner to a Gaussian distribution with a standard deviation of 1 and a mean of 0, facilitating a more balanced and accurate prediction process across all parameters. We note that the hopping amplitude t represents the quantum fluctuation (and consequently also the 2-qubit exchange coupling $J=4t^2/U$), and it is interesting that the machine learning procedure becomes more demanding with increasing quantum fluctuations in the strong-coupling quantum dot qubit problem.



This consideration is largely moot on a practical level, though, since experimentally, a smaller value of $t$ is more realistic \cite{hensgens2017quantum}. With this context, another investigation focuses on the impact on our method when $t$ is small, necessitating high precision in prediction. We previously discussed the 3-dot case, but we also explore the 4 quantum dot scenario with $t \in [0.01,0.25]$, fixing $t_{12}=0.125$ for scale. In this case, an $R^2=0.998$ with errors of $R_{MS}({\delta \epsilon_{i}})=0.0158$, $R_{MS}(\delta V_{i,j})=0.0183$, $R_{MS}(\delta t_{i,j})=0.0155$, and $R_{MS}(\delta U_{i})=0.0442$ was achieved. This demonstrates that even when $\tilde{t}_{ij}$ is small, it can still be predicted accurately for a larger number of quantum dots, and assuming a smaller $t$ increases our method's effectiveness for the other parameters. Interestingly, as we suggested, the ability to predict $\tilde{t}$ is partially limited by the resolution of the charge stability diagrams input into the machine learning scheme, given a fixed amount of training data. This is not surprising, as even without disorder, the ability to determine a fixed $t$ between two quantum dots is limited by the resolution in the charge stability diagram \cite{wang2011quantum}. $t$ affects the diagrams by curving the phase transitions, and discerning smaller values of $t$ requires higher resolution. If a very small $t$ is relevant, utilizing higher resolution charge stability diagrams may be necessary. Moreover, when the range of $t$ is very small, a standard scaling operation can again be applied to the input to prevent the neural network from overlooking the prediction of $t$.

To demonstrate the effects of resolution, we apply our method to a 3 quantum dot system within the restricted $t$ range ($\tilde{t}_{ij} \in [0.01,0.25]$), but with double the resolution—and therefore the number of steps—in the $\mu$ and $\mu_i^j$ parameters. This approach yielded an $R^2=0.9984$ with errors of $R_{MS}(\delta \epsilon_{i})=0.0109$, $R_{MS}(\delta V_{i,j})=0.0160$, $R_{MS}(\delta t_{i,j})=0.0089$, and $R_{MS}(\delta U_{i})=0.0423$. These results mark an improvement in the RMS errors compared to the case with lower resolution. Additionally, employing standard scaling further enhances performance, leading to an $R^2=0.9972$ and $R_{MS}(\delta \epsilon_{i})=0.0110
$, $R_{MS}(\delta V_{i,j})=0.0113$, $R_{MS}(\delta t_{i,j})=0.0045$, and $R_{MS}(\delta U_{i})=0.0347$ { (see Fig. \ref{fig:3AllParamFullSSHighRes})}. Notably, the error $R_{MS}({\delta t_{i,j}})$ is about half that of the lower resolution case without standard scaling. It is is also worth noting that since $R^2$ is affected by standard scaling, it is not directly comparable with the previous results.

A series of tables (I, II, III) with a summary of results for different quantum dot configuration and parameter regimes are given below:

\begin{table}[H]
\begin{tabular}{ c| c c c c c}
 Model & 3$\delta\epsilon$ & 4$\delta\epsilon$ & 4$\delta V_{ij}$ & 4$\delta U_i$ & 4$\delta t_{ij}$ \\
 \hline\hline
 $\# Dots$ & 3 & 4 & 4 & 4 & 4 \\
 \hline
 $R^2$ & 0.99998 & 0.99995 & 0.99996 & 0.99997 & 0.99876\\
 \hline
 Training Data & 45000 & 45808 & 41416 & 41096 & 55672 \\
\hline
 $R_{MS}(\delta \epsilon_i)$ & 0.0018 & 0.0029 & 0 & 0 & 0 \\
\hline
 $R_{MS}(\delta V_{i,j})$ & 0 & 0 & 0.000904 & 0 & 0 \\
 \hline
 $R_{MS}(\delta t_{i,j})$ & 0 & 0 & 0 & 0.02429 & 0\\
 \hline
 $R_{MS}(\delta U_i)$& 0 & 0 & 0 & 0 & 0.10051    
\end{tabular}
\caption{Results from single parameter disorder for different parameters and numbers of quantum dots. The parameter regimes can be found within the main text.}
\end{table}

\begin{table}[H]
\begin{tabular}{ c| c c c c c c}
 Model & 3All & 4All  & 5All\\
 \hline\hline
 $\# Dots$ & 3 & 4 & 5  \\
 \hline
 $R^2$ & 0.9977 & 0.9961 & 0.9949  \\
 \hline
Training Data &  13504 & 13504 & 17475  \\
 \hline
 $R_{MS}(\delta \epsilon_i)$ & 0.0239 & 0.0280 & 0.0321
 \\
\hline
 $R_{MS}(\delta V_{i,j})$ &0.0164& 0.0205& 0.0231  \\
 \hline
 $R_{MS}(\delta t_{i,j})$ & 0.0121 & 0.0193 & 0.0254 \\
 \hline
 $R_{MS}(\delta U_i)$& 0.0373 & 0.0512 & 0.0574    
\end{tabular}
\caption{Results for disorder in all parameters with different numbers of quantum dots. The parameter regimes can be found within the main text.}
\end{table}

\begin{table}[H]
\begin{tabular}{ c| c c c c}
 Model & $3(t<0.25)$ &  $4(t<0.25)$& 3HR (t<0.25) & 3HR(SS)\\
 \hline\hline
 $\# Dots$ & 3 & 4 & 3 & 3 \\
 \hline
 $R^2$ & 0.9987
 & 0.9980
 & 0.9984
 & 0.9972  \\
 \hline
 Training Data &11488
 & 15672
 & 13662 
 & 13662\\
 \hline
 $R_{MS}(\delta \epsilon_i)$ & 0.0128&  0.0158 & 0.0109& 0.0110\\
\hline
 $R_{MS}(\delta V_{i,j})$ & 0.0190 & 0.0183 & 0.0160 & 0.0113\\
 \hline
 $R_{MS}(\delta t_{i,j})$ & 0.0119 & 0.0155 & 0.0089 & 0.0045 \\
 \hline
 $R_{MS}(\delta U_i)$& 0.0358 & 0.0442 & 0.0423 & 0.0347   
\end{tabular}
\caption{Results for disorder in all parameters with a reduced, small magnitude $t\in[0.01,0.25]$ range. 3(t<0.25) and 4(t<0.25) refer to standard resolution training data, while 3HR(t<0.25) and 3HR(SS) refer to the high resolution training data. 3HR(SS) has the additional aspect of being standard scaled as outlined within the paper. Additional details about the parameter regime can be found in the main text.}
\end{table}

\clearpage
\onecolumngrid
\begin{minipage}{\linewidth} 
\begin{figure}[H]
     \centering
     \begin{subfigure}[b]{0.9\linewidth}
         \centering
         \includegraphics[width=\textwidth]{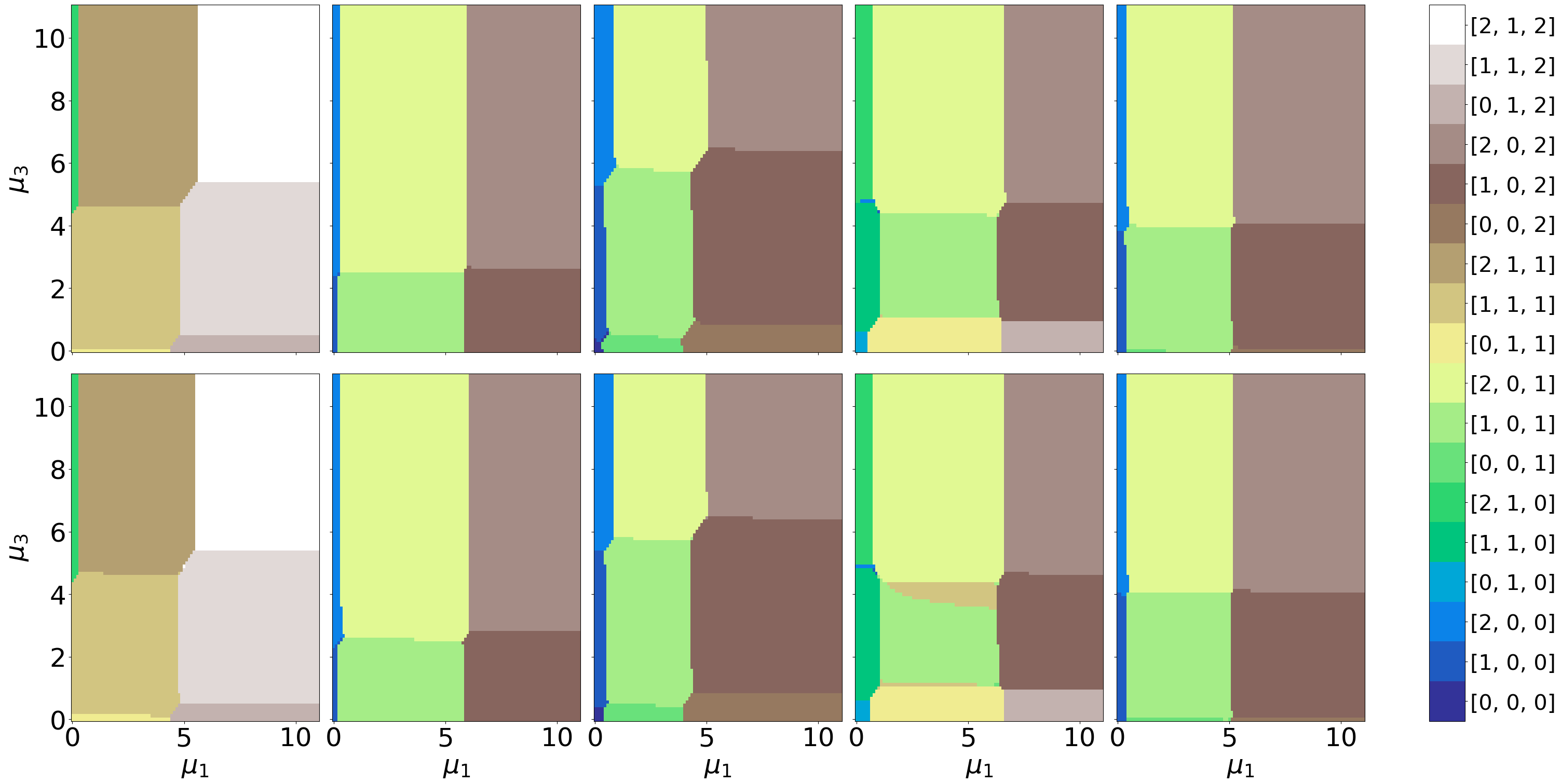}
         \caption{}
     \end{subfigure}
     \begin{subfigure}[b]{0.9\linewidth}
         \centering
         \includegraphics[width=\textwidth]{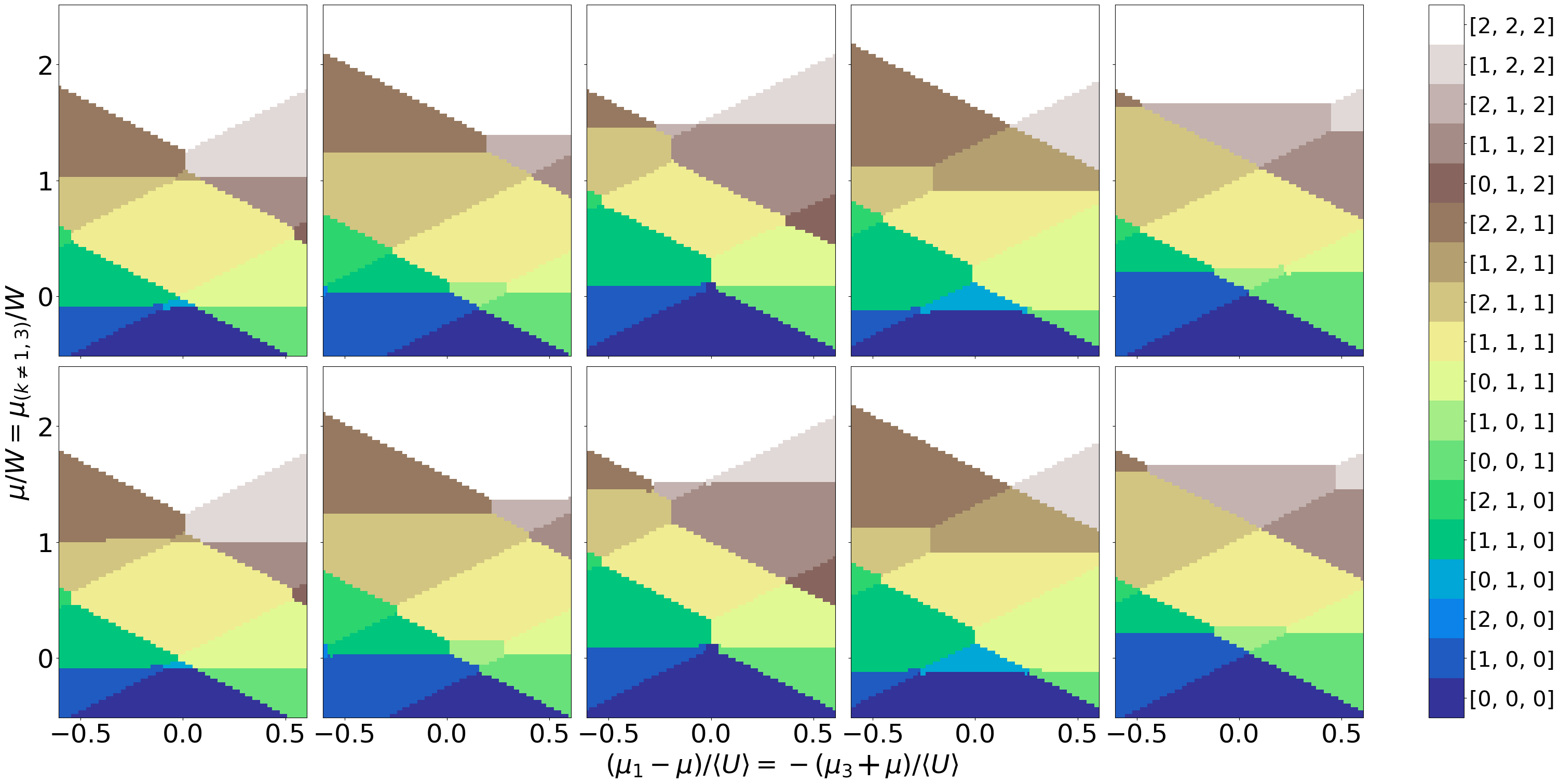}
         \caption{}
     \end{subfigure}
    \caption{3 quantum dot charge stability diagrams for input and expected measurement outcomes from Hubbard model parameters for the prediction of disorder in all parameters with a reduced $t \in [0.1,0.25]$ range. The root mean squared error in the Hubbard model parameters was $R_{MS}({\delta \epsilon_{i}})=0.0128$, $R_{MS}({\delta V_{i,j}})=0.0190$, $R_{MS}({\delta t_{i,j}})=0.0119$, $R_{MS}({\delta U_{i}})=0.0358$ with an $R^2=0.9987$. In both plots the first row is the input charge stability diagrams, namely the most probable state for a chemical potential configuration, and the second row is the expected charge stability diagram given the prediction of the Hubbard model parameters. (a) Most probable state for input Hubbard parameters (1st row) and predicted Hubbard parameters (2nd row) where $\mu_1$ and $\mu_3$ are independently varied. (b) Most probable state for input Hubbard parameters (1st row) and predicted Hubbard parameters (2nd row) where the chemical potential at each site is $\vec{\mu}=[\mu_1,\cdots,\mu_n]$ with our plot having axis $\mu_1=-\mu_3$ (rescaled by $\langle U \rangle$ and shifted by $\mu$) vs. $\mu_{k\neq 1,3}=\mu$ (rescaled by $W=\frac{1}{L}\left(\sum_i \langle U_i \rangle +\langle V_{i,i+1}\rangle\right)$ where $\langle U\rangle=4$ and $\langle V_{i,i+1}\rangle=0.2$ are the their disorder free values respectively ), this is the data input into the machine learning model. 
    }     \label{fig:3AllParamFullS}
\end{figure}
\end{minipage}

\begin{minipage}{\linewidth} 

\vspace{-0.1cm}
\end{minipage}

\begin{minipage}{\linewidth} 
\begin{figure}[H]
     \centering
     \begin{subfigure}[b]{0.9\linewidth}
         \centering
         \includegraphics[width=\textwidth]{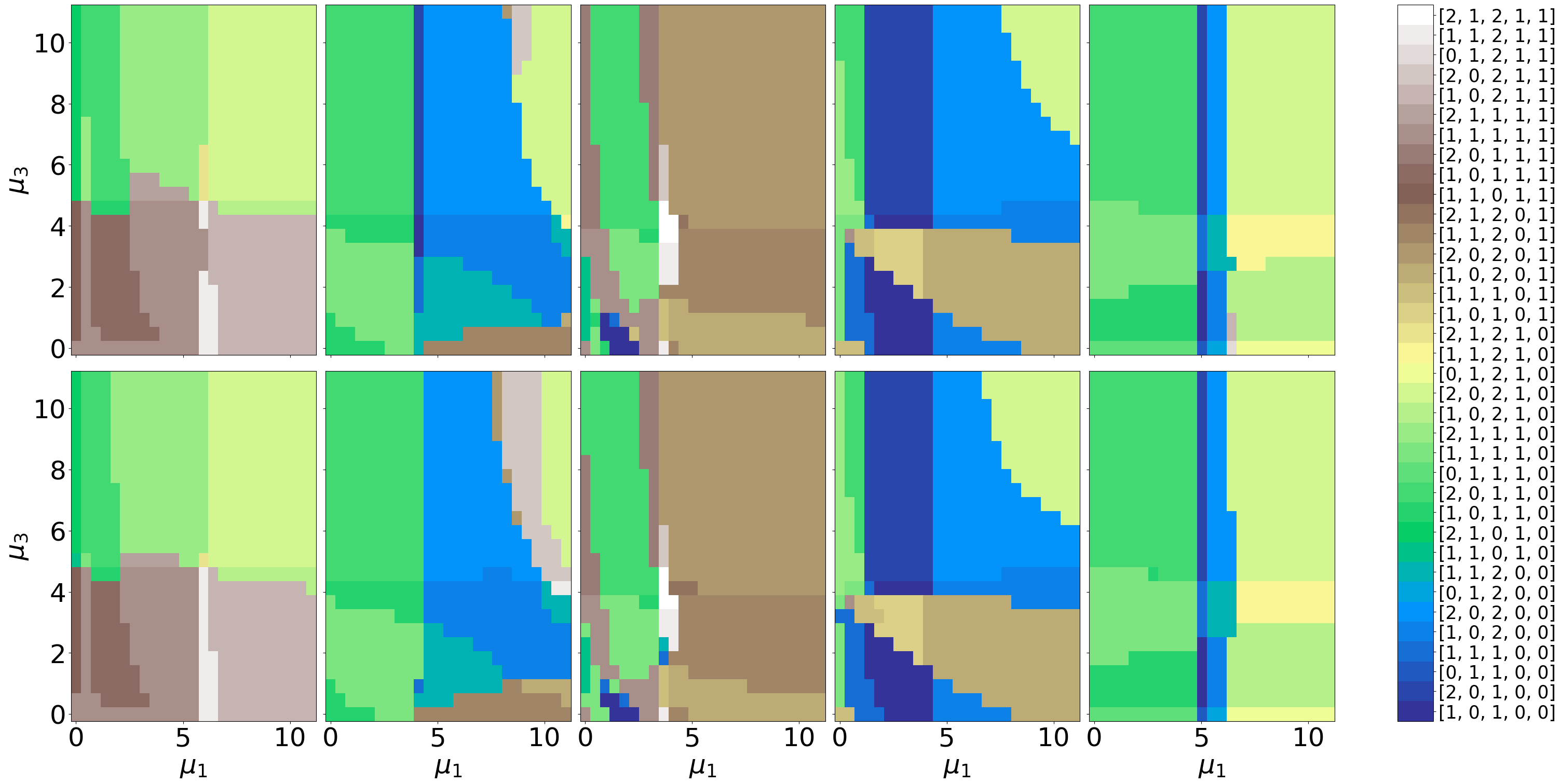}
         \caption{}
     \end{subfigure}
     \begin{subfigure}[b]{0.9\linewidth}
         \centering
         \includegraphics[width=\textwidth]{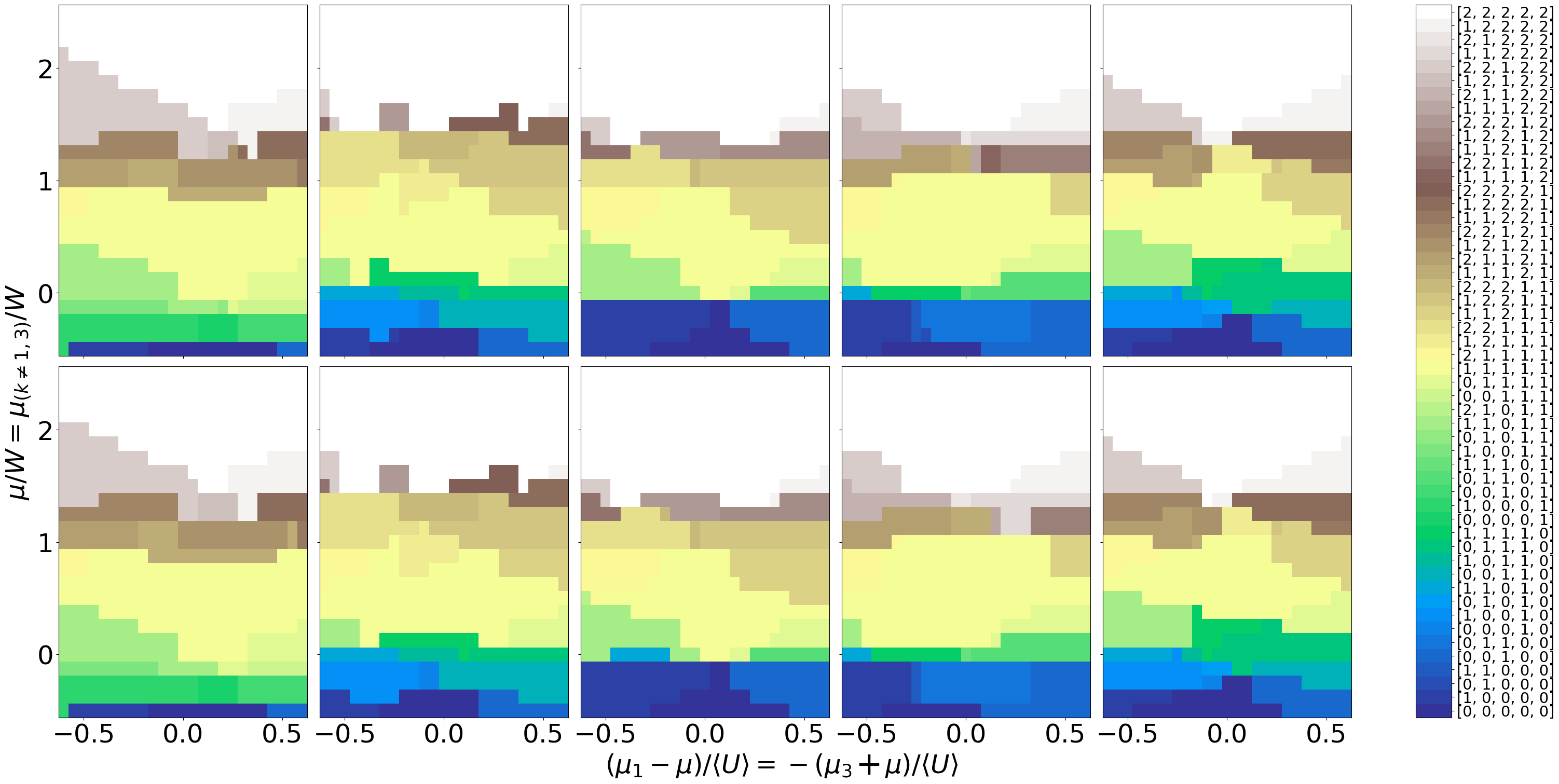}
         \caption{}
     \end{subfigure}

    \caption{5 quantum dot charge stability diagrams for input and expected measurement outcomes from Hubbard model parameters for the prediction of disorder in all parameters. The root mean squared error in the Hubbard model parameter deviations was $R_{MS}(\delta \epsilon_{i})=0.0321$, $R_{MS}(\delta V_{i,j})=0.0231$, $R_{MS}(\delta t_{i,j})=0.0254$, $R_{MS}(\delta U_{i})=0.0574$ with an $R^2=0.995$. In both plots the first row is the input charge stability diagrams, namely the most probable state for a chemical potential configuration, and the second row is the expected charge stability diagram given the prediction of the Hubbard model parameters. (a) Most probable state for input Hubbard parameters (1st row) and predicted Hubbard parameters (2nd row) where $\mu_1$ and $\mu_3$ are independently varied. (b) Most probable state for input Hubbard parameters (1st row) and predicted Hubbard parameters (2nd row) where the chemical potential at each site is $\vec{\mu}=[\mu_1,\cdots,\mu_n]$ with our plot having axis $\mu_1=-\mu_3$ (rescaled by $\langle U \rangle=4$ and shifted by $\mu$) vs. $\mu_{k\neq 1,3}=\mu$ (rescaled by $W=\frac{1}{L}\left(\sum_i \langle U_i \rangle +\langle V_{i,i+1}\rangle\right)$ where $\langle U\rangle=4$ and $\langle V_{i,i+1}\rangle=0.2$ are the their disorder free values respectively ), this is similar to the stability diagrams fed into the neural network except the neural network only receives $\mu_{i}=-\mu_j$ between nearest neighbors $i=j+1$.
    }     \label{fig:5AllParamFull}

\end{figure}
\end{minipage}
\begin{minipage}{\linewidth}
\begin{figure}[H]
     \centering
     \begin{subfigure}[b]{0.65\linewidth}
         \centering
         \includegraphics[width=\textwidth]{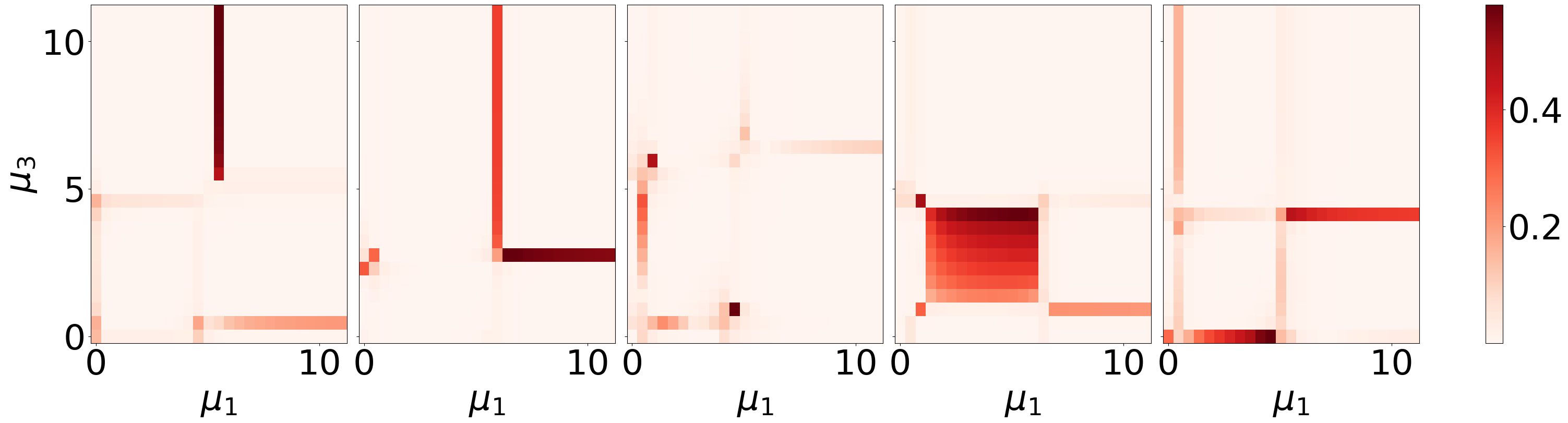}
         \caption{}
     \end{subfigure}
     \begin{subfigure}[b]{0.65\linewidth}
         \centering
         \includegraphics[width=\textwidth]{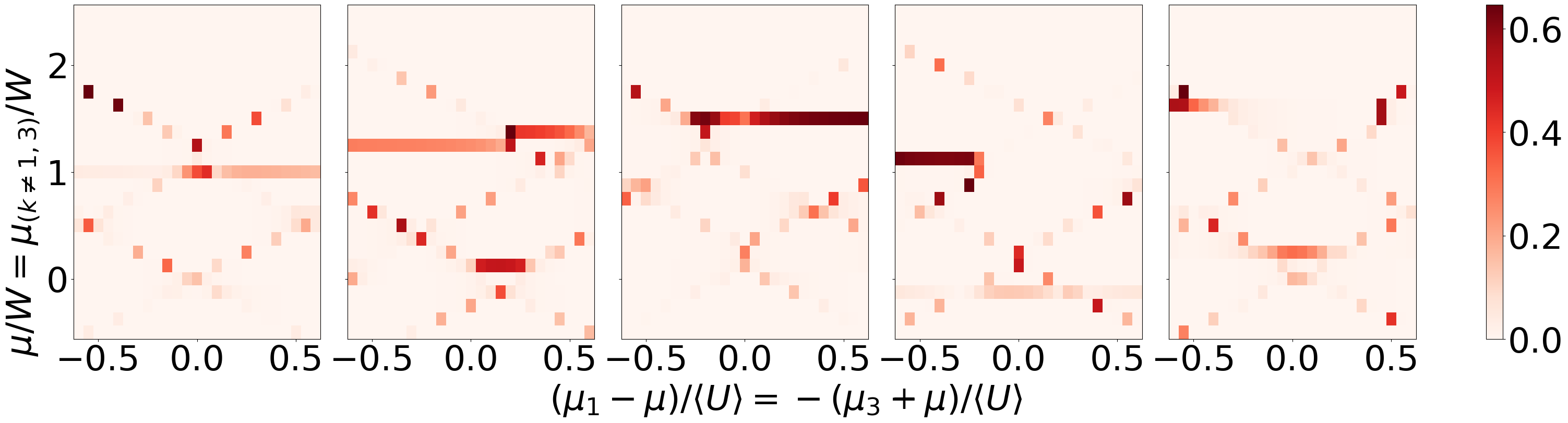}
         \caption{}
     \end{subfigure}
\vspace{-0.2cm}
    \caption{Difference in 3 quantum dot in charge stability diagrams for input and expected measurement outcomes from Hubbard model parameters for the prediction of disorder in all parameters with a reduced $t \in [0.1,0.25]$ range. In particular $n_{error}=||\langle\vec{n}_{input}\rangle-\langle\vec{n}_{predicted}\rangle||$ is plotted in both plots where $\langle\vec{n}_{input}\rangle$ is the occupation expectation vector for all sites of the input data and $\langle\vec{n}_{predicted}\rangle$ is the expected occupation expectation vector given the prediction of the Hubbard model parameters. The columns of both plots correspond to the charge stability diagrams in Fig. \ref{fig:3AllParamFullS} with the same column. The error-free model parameters were set to $U=4$, $t=0.125$ and $V_{i,i+1}=0.2$. (a) $n_{error}$ between input Hubbard parameters and predicted Hubbard parameters where $\mu_1$ and $\mu_3$ are independently varied. (b) $n_{error}$ between input Hubbard parameters and predicted Hubbard parameters where the chemical potential at each site is $\vec{\mu}=[\mu_1,\cdots,\mu_n]$ with our plot having axis $\mu_1=-\mu_3$ (rescaled by $\langle U \rangle$ and shifted by $\mu$) vs. $\mu_{k\neq 1,3}=\mu$ (rescaled by $W=\frac{1}{L}\left(\sum_i \langle U_i \rangle +\langle V_{i,i+1}\rangle\right)$ where $\langle U\rangle=4$ and $\langle V_{i,i+1}\rangle=0.2$ are the their disorder free values respectively).}
     \label{fig:E3AllParamFullS}
\end{figure}
\vspace{-0.1cm}
\begin{figure}[H]
     \centering
     \begin{subfigure}[b]{0.65\linewidth}
         \centering
         \includegraphics[width=\textwidth]{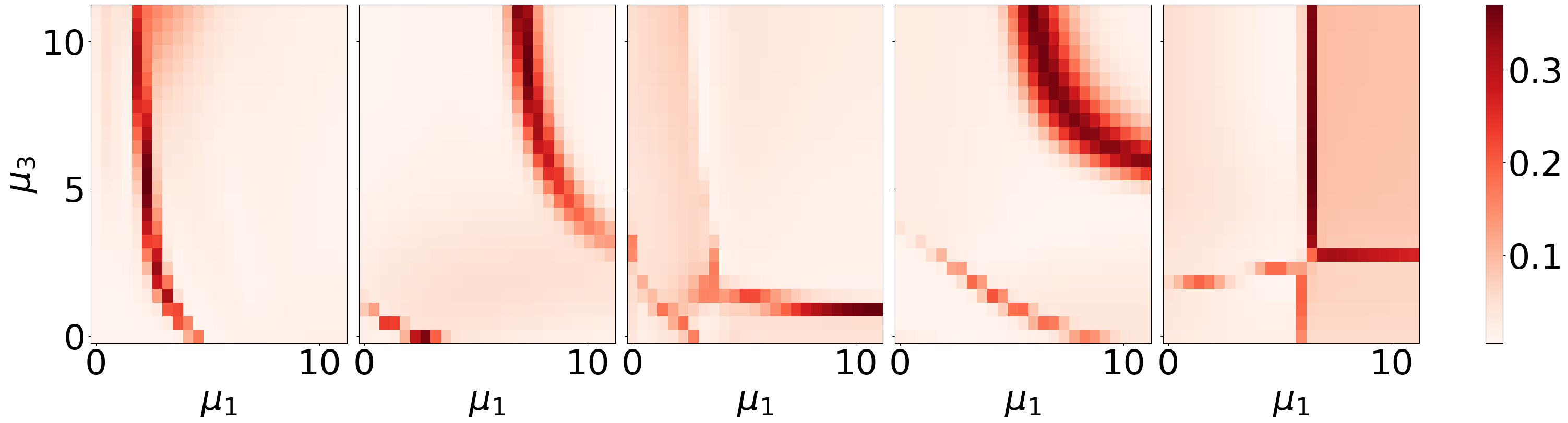}
         \caption{}
     \end{subfigure}
     \begin{subfigure}[b]{0.65\linewidth}
         \centering
         \includegraphics[width=\textwidth]{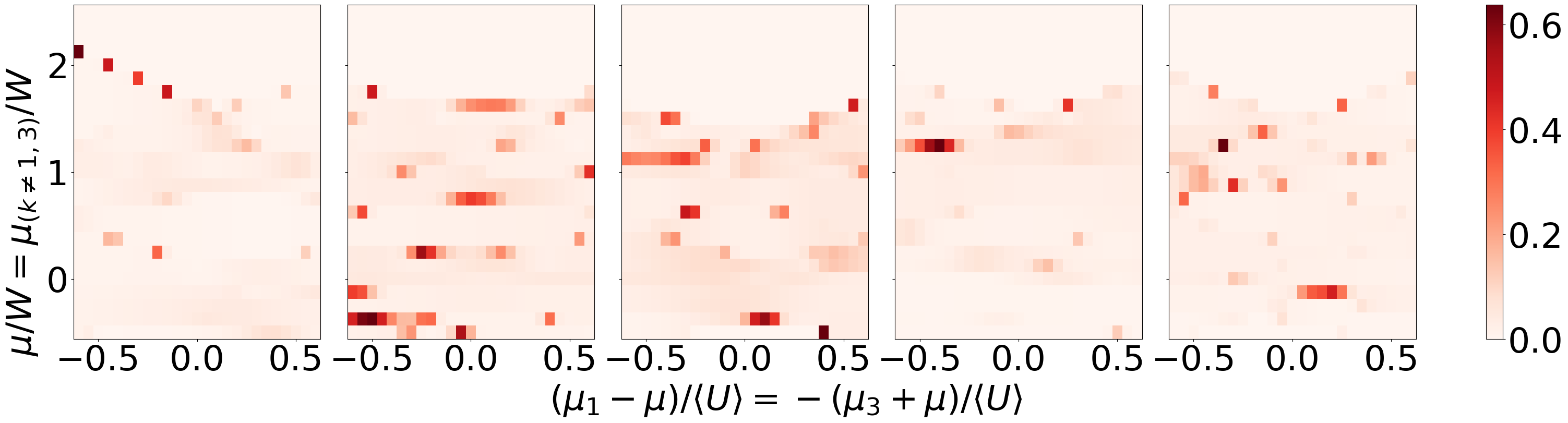}
         \caption{}
     \end{subfigure}
     \vspace{-0.2cm}
    \caption{Difference in 5 quantum dot in charge stability diagrams between input and expected measurement outcomes from Hubbard model parameters for the prediction of disorder in all parameters. In particular $n_{error}=||\langle\vec{n}_{input}\rangle-\langle\vec{n}_{predicted}\rangle||$ is plotted in both plots where $\langle\vec{n}_{input}\rangle$ is the occupation expectation vector for all sites of the input data and $\langle\vec{n}_{predicted}\rangle$ is the expected occupation expectation vector given the prediction of the Hubbard model parameters. The columns of both plots correspond to the charge stability diagrams in Fig. \ref{fig:5AllParamFull} with the same column. The error-free model parameters were set to $U=4$, $t=1$ and $V_{i,i+1}=0.2$. (a) $n_{error}$ between input Hubbard parameters and predicted Hubbard parameters where $\mu_1$ and $\mu_3$ are independently varied. (b) $n_{error}$ between input Hubbard parameters and predicted Hubbard parameters where the chemical potential at each site is $\vec{\mu}=[\mu_1,\cdots,\mu_n]$ with our plot having axis $\mu_1=-\mu_3$ (rescaled by $\langle U \rangle=4$ and shifted by $\mu$) vs. $\mu_{k\neq 1,3}=\mu$ (rescaled by $W=\frac{1}{L}\left(\sum_i \langle U_i \rangle +\langle V_{i,i+1}\rangle\right)$ where $\langle U\rangle=4$ and $\langle V_{i,i+1}\rangle=0.2$ are the their disorder free values respectively).}\label{fig:E5AllParamFull}
\end{figure}
\end{minipage}

\clearpage
\twocolumngrid
\subsection{No Trust Verification { Scheme}}

One of the strengths of this method is that if one can accurately determine the Hubbard model parameters, then almost any measurement one can perform on the quantum dot qubits could also be simulated. This implies (as will be discussed) that one does not need to blindly trust the machine learning algorithm at all to be confident in the results it provides. In this regard, one could create a new $K'$ with a completely different configuration of experimental parameters such as:

$$K'=[\mu_1^j,\cdots\mu_N^j, T^j]$$

This particular configuration of experimental parameters could allow the site-dependent chemical potentials to be set to any value within a reasonable range and even incorporate variation of the temperature ($T$). One could even add new measurement types. The Hubbard model, and thus a generator function, is capable of simulating any of these possible $K'$ configurations. Inserting the new $K'$ into our generator function along with the prediction of the Hubbard model parameters from the CNN, $Y$, would yield a new measurement matrix, $X'$.

$$f_{gen}(Y,K')=X'$$

Here, we { propose to} simulate performing measurements in these different configurations (including new measurements beyond occupation numbers) to generate a new $X'$. Since one could then carry out those measurements experimentally for the $K'$ measurement configurations and compare their outcomes to that of $X'$, a match would serve as strong evidence of the validity of the disorder (or, more specifically, Hubbard model parameter) predictions. One could generate an arbitrarily large set of new measurement configurations, $K'$, to test the predictions, and all reasonable $K$ matrices should still match in both simulated and experimental results if the $Y$ prediction is accurate. Since this new $K'$ never enters the machine learning scheme, the neural network would be incapable of generating Hubbard parameters that "hallucinate" a match with these measurements. { Since the probability of a "hallucination" result passing these additional verification checks is low, it would likely not be necessary to perform many additional measurements.} Thus, without any trust of the machine learning scheme, one could perform as many measurements as necessary to ensure that the Hubbard model parameters predicted by the algorithm are correct. { These verifications also may allow additional experimental tuning of the NN in the cases of failure, for instance minor modifications to the model Hamiltonian, a higher resolution in the charge stability diagram or merely more training data could be required to achieve accuracy for some experimental applications, the verification scheme can work to do this without being mislead by potential "hallucinations".}

\subsection{Costs}
\label{S:Cost}
Let us consider how this method would be applied to a large number of quantum dots. It is important to note that our method cannot be scaled up indefinitely, as eventually, the exponential scaling of the Hubbard model simulations will render it unfeasible. (This has nothing whatsoever to do with any limitation in our machine learning method or algorithm, it is simply that generating Hubbard model exact diagonalization solutions is an exponentially hard problem, which becomes unfeasible for more than {15}-20 quantum dots because of the exponential growth of the Hilbert space.) However, it does permit simultaneous tuning of many more dots in aggregate at once than manual tuning allows. It is crucial to emphasize that the machine learning aspect is not the limiting factor; in fact, we find that the neural network and the amount of training data required do not seem to scale significantly with the addition of more quantum dots. The primary limitation is the exponential scaling of the Hubbard model used in generating the training data, which restricts our method's scalability to an arbitrary number of quantum dots. {\comment{The fact that the machine learning numerics itself do not scale with the number of dots indicates that perhaps our technique may be useful in quantum computing platforms with hundreds of quantum dot qubits since automating the control may involve only 2-5 dots at a time, as discussed below.}}

\begin{figure}[H]
     \centering
     \begin{subfigure}[b]{0.9\linewidth}
         \centering
         \includegraphics[width=\textwidth]{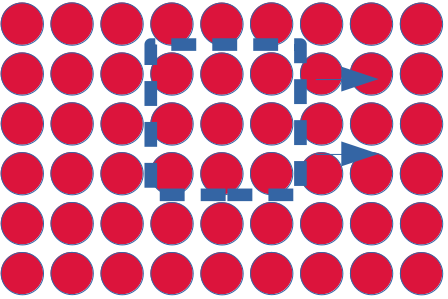}
         \caption{}
     \end{subfigure}
    \caption{Diagram of the shifting local tuning process: For large systems, a local window, illustrated by the blue rectangle, is shifted from site to site to tune the dot within the center. The neural network would be trained only once on the local system equal to the size of the window and used to determine Hubbard deviations between the center dot and its local neighbors.}     \label{fig:WindowPlot}
\end{figure}

The notion of scaling up to tune an arbitrary number of quantum dots simultaneously is, to begin with, excessive and unnecessary. The necessity for tuning quantum dots simultaneously is not dictated by the total number of dots within the system but by the locality of the system. For a large system with, say, 100s or more quantum dots, the envisioned tuning process involves adjusting a quantum dot within a locality that includes a certain number of nearest and next-nearest neighbors dependent on the interaction range. { The effect another dot has on the central dots model parameters decreases the further away it is.} In this localized window within the bulk, one would determine the parameters of the central quantum dot(s) and then shift the window through the system, tuning a limited number of quantum dots at a time while considering the effects of a broader locality on the charge stability diagram. It is worth mentioning that as the window shifts, a new neural network would not need to be trained; one could simply reuse the charge stability diagrams measured from the initial locality for subsequent tunings. Thus, our technique may provide the pathway for the automated control of Hubbard model based quantum dot platforms of the future containing arbitrarily large number of qubits.

The ability of our method to tune more quantum dots simultaneously becomes particularly crucial when considering 2D geometries. In a 2D grid of quantum dots, current methods that tune two or three quantum dots at the same time are likely to face challenges, as each dot would have four nearest neighbors, even if we only consider nearest neighbors as relevant. In such a case even these nearest neighbors would significantly influence the charge stability diagram between any two dots. Our method, in principle, could be adapted to tackle such scenarios, enabling the tuning of much more complex geometries through the use of a specified locality window. 
{ Moreover, in terms of linear quantum dot chains, we suspect 5 dots may be sufficient for most near-term devices since using current manual methods chains of length 12 are able to be tuned by 2-3 dots at a time \cite{neyens2024probing}. In terms of scaling to 100s of dots with complex geometries, while the locality required is not known and thus our method is not guaranteed to work for arbitrarily large arrays, it is a potentially viable path for which no alternative exists.}
{\comment{Moreover, since current systems utilize quantum dot chains, it appears that tuning up to 5 quantum dots at a time, as we currently achieve, would be sufficient for a system with any number of dots since the interactions are likely to be local and limited to a few (<5) dots only.}}

{The scalability of our approach is not bottlenecked by the CNN, but hinges on our ability to generate training data, specifically how many quantum dots can be simulated at once.} The Hubbard model operates within a strongly interacting regime, setting a limit on the number of dots that can be simulated. For us, it takes 30 minutes for one core to generate a single training realization for 5 quantum dots (within the $\tilde{t}\in[0.1,2.0]$ regime); however, our training data generation is highly inefficient, as it does not utilize any symmetries and relies solely on direct exact diagonalization. Adopting more advanced methods \cite{qin2022hubbard} would enable the simulation of significantly more quantum dots simultaneously (perhaps up to 20-30). Moreover, since the generation process is parallelizable for each row of the K matrix, the resources required effectively comes down to the number of cores one has access to. We do not observe significant scaling in the number of training data realizations required for tuning quantum dots; hence, for 5 quantum dots and 15,000 training realizations, about 7.5 kilo-core hours are needed. A significant speedup is achievable if thermal effects are disregarded and only the ground state is considered (which seems reasonable at very low temperatures within the parameter range \cite{hensgens2017quantum}), reducing the time for a single training realization for 5 quantum dots to about 5 minutes. Assuming the inefficient case of $O(4^N)$ scaling, our resources and (even our inefficient diagonalization) method would necessitate a large but feasible 1920 kilo-core hours for tuning 9 quantum dots. If focusing solely on the ground state, even with our inefficient methods, computing 9 quantum dots would require only about 320 kilo-core hours. In the scenario of a 2D square grid, accounting for nearest and next-nearest neighbor interactions would involve 3x3=9 quantum dots. Given these considerations, we see no reason why this could not be achieved using our method. This becomes even more feasible when targeting only the ground state or a few excited states, as such an approach would enable the use of tensor network methods such as DMRG or PEPS, which have been used to find the ground state of much larger (up to 8x8) and similar Hubbard model systems \cite{stoudenmire2012studying,scheb2023finite}.

\subsection{Generalizing to Other Problems}
Our proposed method of addressing aggregate invertible disorder problems through machine learning can be and has been extended to other problems. In our previous work, we demonstrated the feasibility of inverting the aggregate conductances of a Majorana nanowire \cite{taylor2023machine} to determine its disorder, providing two distinct cases where this scheme was utilized to determine the disorder. In contrast to the current Hubbard model problem, the Majorana problem is essentially a single-particle problem and can therefore be exactly diagonalized for very large systems.

In both cases, by performing an aggregation of measurements from a physical system, one can determine an unknown disorder within the system. This aspect of aggregation is quite important because machine learning can uniquely solve for the unknown disorder, sometimes even in cases where individual measurement are not invertible. For instance, in the case of the Majorana nanowire, the scattering problem is not in general known to be invertible; yet, despite this, the aggregate of many scattering problems seems to be invertible \cite{taylor2023machine}. Similarly, it is clear that a single measurement of occupation expectations is not sufficient in the current quantum dot system to determine the deviation in Hubbard parameters; however, a sequence of such measurements, in the form of many charge stability diagrams, does seem to be uniquely invertible. By restricting one's range of parameters to those that are experimentally relevant and performing an aggregation of measurements of such devices, it is possible to uniquely determine those disorder parameters. Furthermore, it seems that neural networks are able to do so efficiently.

The "no trust" component can also be generalized to many problems, provided there is a usable generator function. This represents a significant improvement over previous results because neural networks are prone to hallucination, and without this sanity check, one cannot generally trust that a neural network's output is valid in experimental cases. However, with this sanity check, one can confirm the machine learning results with arbitrary confidence, thereby making this method an extremely viable tool for future research.

\section{Conclusion}

In conclusion, we provide a method of solving disorder landscapes though a series of aggregate measurements that could be generalized to many physical models. In this work we also {propose} a novel "no trust" verification method that virtually eliminates the risk of "hallucination" predictions by the neural network. This provides a method of conclusively verifying the predictions of the neural network such that even without any trust of the neural network one can be confident about the results.

{Our technique explicitly uses deep learning on the whole system together and is not focused on tuning small parts (e.g. a few dots in a many quantum dot system) of the system individually using automated control, making our method particularly useful and powerful.} This work successfully demonstrates a method, which by employing machine learning techniques, specifically through the use of a CNN, is able to accurately determine disordered deviations in the parameters of the extended Hubbard model as {applied} to quantum dot qubits. By analyzing a series of charge stability diagrams from nearest neighbor pairs of quantum dots, our CNN is adept at identifying spatially dependent disorder across each parameter of the model. This approach shows remarkable proficiency in predicting with high accuracy the spatially dependent variations in coupling constants, both Coulomb intra and inter-site terms, as well as site-specific gate voltage errors. These results are of {\comment{particular}} experimental interest due to the natural description of coupled semiconductor quantum dots through the Hubbard model and the readily experimental availability of charge stability diagram measurements. Although we discuss our results in terms of disorder, our technique in fact is a method to obtain the extended Hubbard model Hamiltonian parameters themselves using just the stability diagrams as the input.  Even if all the parameters underlying the quantum dot qubit platform are unknown, our technique, in principle, can provide an accurate estimate for all the parameters of the underlying Hamiltonian as long as sufficient amount of input charge stability diagrams are available. {This is without the need for any additional information on the system, beyond some rough (potentially order of magnitude) estimate on what values the Hubbard parameters can take for training data generation. We emphasize, it is not necessary to get an initial site specific estimate of any Hubbard model parameter, a vague all sites have $U\in[0,8]$ for instance is sufficient to generate training data and predict any particular sites $\tilde{U}_i$.} Thus, the technique developed in this work is not only applicable to semiconductor quantum dot qubit platforms, but also to the strongly correlated Hubbard model emulations.

Extending beyond previous studies, our investigation includes the prediction of disorder not only in individual parameters per site but also across all parameters simultaneously. This provides a comprehensive {prediction} of disorder within the system, marking a notable enhancement over earlier methods that focused primarily on single parameter deviations, such as gate voltages \cite{hensgens2017quantum}. Additionally, our research broadens the scope of conventional quantum dot tuning practices by considering the interactions among 3, 4, and 5 quantum dots simultaneously, thereby integrating crosstalk effects that are typically disregarded in the standard methodology of tuning quantum dot pairs.

{\comment{This study is the first to solve site-specific disordered parameters within the extended Hubbard model for configurations exceeding two quantum dots, covering aspects such as coupling constants, on-site potentials, and inter-site repulsions, and extending to site-specific disordered gate voltages for more than three quantum dots. The introduction of our CNN not only paves the way for potentially fully automated tuning of quantum dot-based systems, eliminating the laborious manual adjustment process when parameters drift, but also ensures the reliability of the tuning process through a "no trust" verification method.}}

We mention that in quantum dot qubits {\comment{(and in fact, in all quantum computing platforms),}} the background disorder drifts in the sample changing the disorder landscape slowly over time,  creating a charge noise.  Typically, these drifts are slow and the noise is quasistatic, which is what our model for disorder assumes. {This quasistatic approximation for the charge disorder is consistent with the fact that the associated noise manifests the well-known '1/f frequency spectrum' indicating that the noise s maximized at low frequencies (i.e. small values of f), where of course the quasistatic approximation is essentially exact.}  But such a drift{, even if it is quasistatic,} of course causes the practical problem that an already perfectly tuned experimental quantum dot system drifts out of its tuned parameter regime over time, and must be retuned.  This is the key problem.  Our machine learning technique developed in this work, when embedded into the system, { if successful, }would enable the retuning in an automated way without manual human intervention since all that is needed is a measurement of the charge stability diagram allowing the system to be retuned, which can be incorporated into the measurement and tuning electronics. {The only necessary ingredient is the training of the algorithm using a sufficient amount of input stability diagrams obtained from measurements on the quantum dot circuits, as established in the current work using our simulations with different background disorder (which simulates the effective drift and noise).}

Further, though the use of higher resolution charge stability diagram input, our investigation seems to show that for smaller coupling constants the accuracy in predictions of the disorder can be increased as as much as necessary to fit the needs of experimentalists. We also show that a smaller $t_{ij}$ tends to increase the fidelity of the results in the other parameters. Additionally, our results show that the method continues to work as the number of quantum dots increases with little to no decrease in fidelity. While we do not envision using our method for a whole computational device (due to the scaling of computation in generating the training data), our technique {likely} could be scaled up to match the locality of a device to tune many quantum dots at once.

Looking to the future,  while our method is {expected to be} able to apply directly to experiment and {\comment{significantly}} improve the present tuning process of quantum dot devices, it should be possible to extend it to start from a microscopic Hamiltonian to determine directly the semiconductor quantum dot parameters. In fact, there is no reason to think that the technique itself is limited only to the Hubbard model, and in principle, it can be used to study other models of quantum dot qubits effectively. Our current use of the extended Hubbard model is based only on the fact that such a model seems to describe the semiconductor quantum dot based qubit platforms reasonably well.


\section{Acknowledgement}
This work is supported by the Laboratory for Physical Sciences.

\bibliography{mainbib}

\section{Appendix: Additional Figures}
Within the following pages we provide additional figures referenced in the text.

\setcounter{figure}{0}
\renewcommand{\thefigure}{A\arabic{figure}}
{
\newpage
\clearpage
\onecolumngrid
\begin{minipage}{\linewidth} 
\begin{figure}[H]
     \centering
     \begin{subfigure}[b]{0.9\linewidth}
         \centering
         \includegraphics[width=\textwidth]{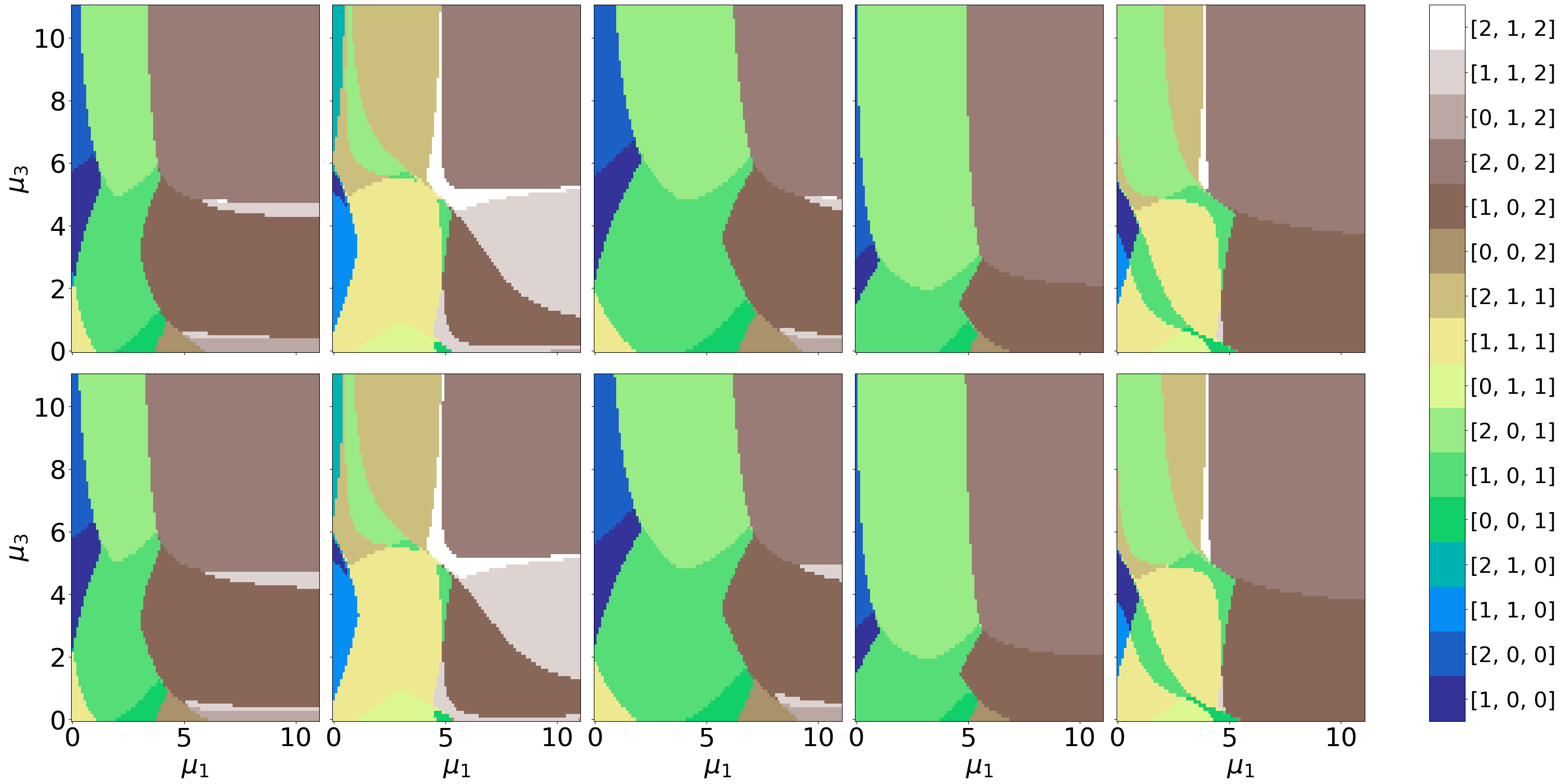}
         \caption{}
     \end{subfigure}
     \begin{subfigure}[b]{0.9\linewidth}
         \centering
         \includegraphics[width=\textwidth]{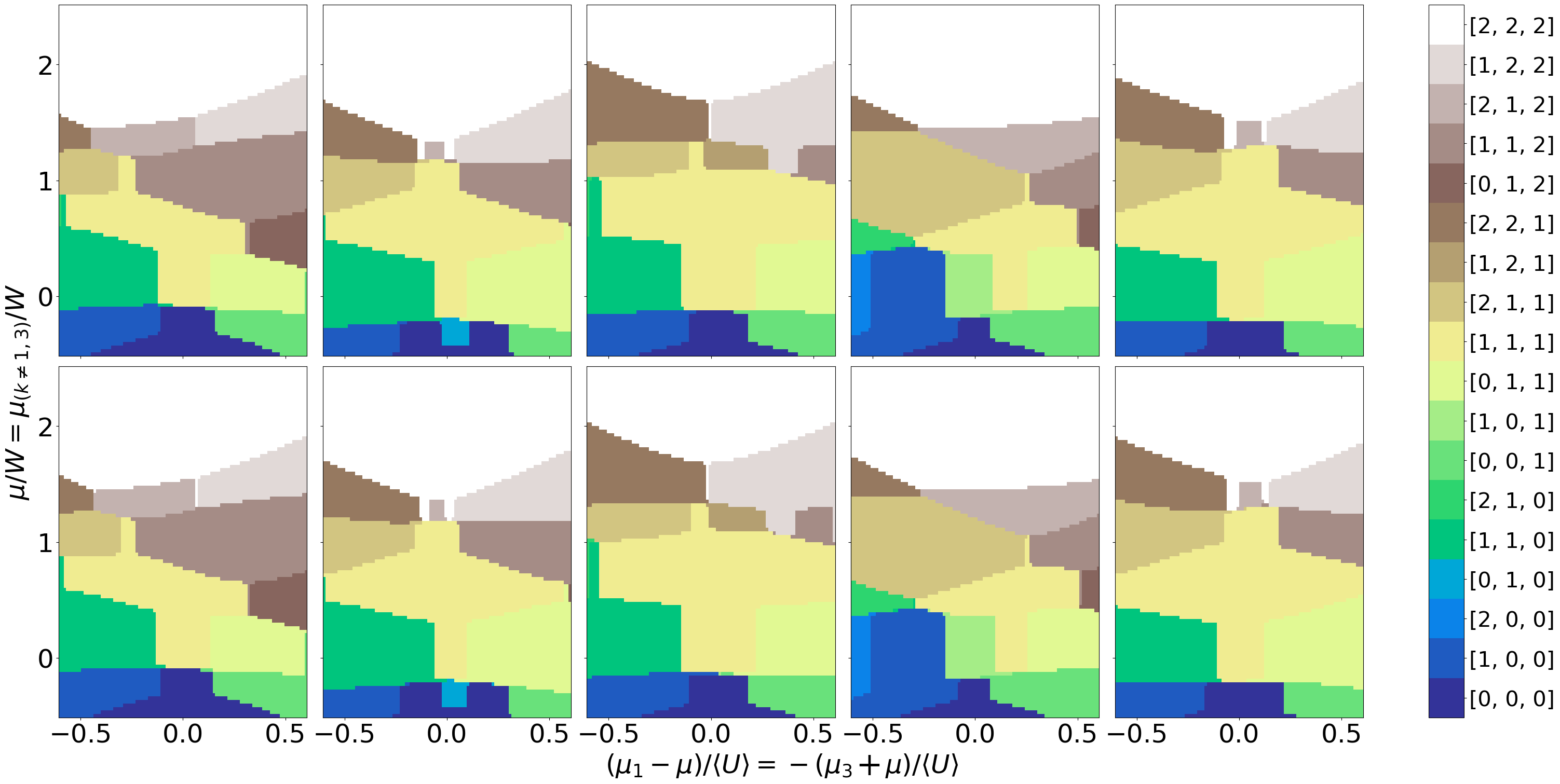}
         \caption{}
     \end{subfigure}

    \caption{{3 quantum dot charge stability diagrams for input and expected measurement outcomes from Hubbard model parameters for the prediction of disorder in all parameters. The root mean squared error in the Hubbard model parameters was $R_{MS}({\delta \epsilon_{i}})=0.0239$, $R_{MS}({\delta V_{i,j}})=0.0164$, $R_{MS}({\delta t_{i,j}})=0.0121$, $R_{MS}({\delta U_{i}})=0.0373$ with an $R^2=0.9977$. In both plots the first row is the input charge stability diagrams, namely the most probable state for a chemical potential configuration, and the second row is the expected charge stability diagram given the prediction of the Hubbard model parameters. (a) Most probable state for input Hubbard parameters (1st row) and predicted Hubbard parameters (2nd row) where $\mu_1$ and $\mu_3$ are independently varied. (b) Most probable state for input Hubbard parameters (1st row) and predicted Hubbard parameters (2nd row) where the chemical potential at each site is $\vec{\mu}=[\mu_1,\cdots,\mu_n]$ with our plot having axis $\mu_1=-\mu_3$ (rescaled by $\langle U \rangle$ and shifted by $\mu$) vs. $\mu_{k\neq 1,3}=\mu$ (rescaled by $W=\frac{1}{L}\left(\sum_i \langle U_i \rangle +\langle V_{i,i+1}\rangle\right)$ where $\langle U\rangle=4$ and $\langle V_{i,i+1}\rangle=0.2$ are the their disorder free values respectively), this is the data input into the machine learning model. 
    }}\label{fig:3AllParamFull}
\end{figure}
\end{minipage}
\clearpage
\onecolumngrid
\begin{minipage}{\linewidth} 
\begin{figure}[H]
     \centering
     \begin{subfigure}[b]{0.65\linewidth}
         \centering
         \includegraphics[width=\textwidth]{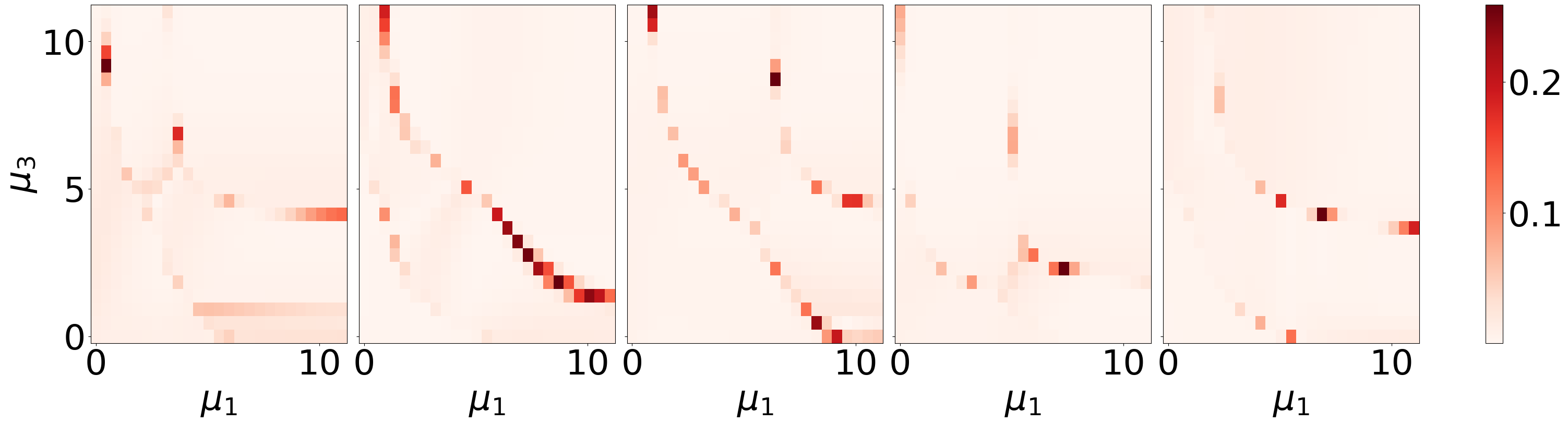}
         \caption{}
     \end{subfigure}
     \begin{subfigure}[b]{0.65\linewidth}
         \centering
         \includegraphics[width=\textwidth]{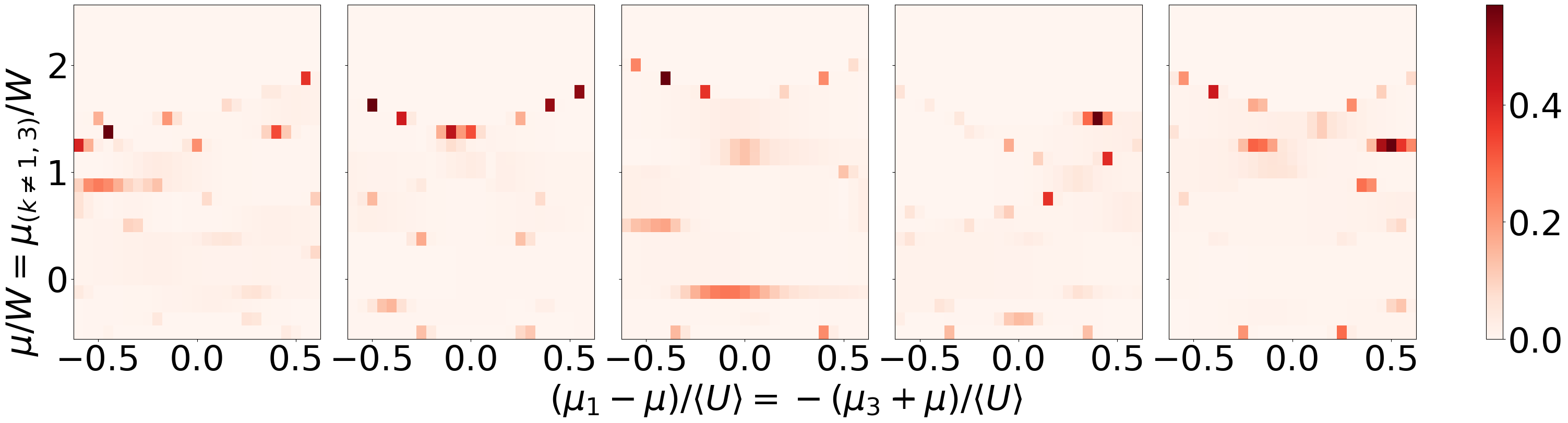}
         \caption{}
     \end{subfigure}
\vspace{-0.2cm}
    \caption{{Difference in 3 quantum dot in charge stability diagrams for input and expected measurement outcomes from Hubbard model parameters for the prediction of disorder in all parameters. In particular $n_{error}=||\langle\vec{n}_{input}\rangle-\langle\vec{n}_{predicted}\rangle||$ is plotted in both plots where $\langle\vec{n}_{input}\rangle$ is the occupation expectation vector for all sites of the input data and $\langle\vec{n}_{predicted}\rangle$ is the expected occupation expectation vector given the prediction of the Hubbard model parameters. The columns of both plots correspond to the charge stability diagrams in Fig. \ref{fig:3AllParamFull} with the same column. The error-free model parameters were set to $U=4$, $t=1$ and $V_{i,i+1}=0.2$. (a) $n_{error}$ between input Hubbard parameters and predicted Hubbard parameters where $\mu_1$ and $\mu_3$ are independently varied. (b) $n_{error}$ between input Hubbard parameters and predicted Hubbard parameters where the chemical potential at each site is $\vec{\mu}=[\mu_1,\cdots,\mu_n]$ with our plot having axis $\mu_1=-\mu_3$ (rescaled by $\langle U \rangle$ and shifted by $\mu$) vs. $\mu_{k\neq 1,3}=\mu$ (rescaled by $W=\frac{1}{L}\left(\sum_i \langle U_i \rangle +\langle V_{i,i+1}\rangle\right)$ where $\langle U\rangle=4$ and $\langle V_{i,i+1}\rangle=0.2$ are the their disorder free values respectively).}}
         \label{fig:E3AllParamFull}
\end{figure}
\vspace{-0.2cm}

\begin{figure}[H]
     \centering
     \begin{subfigure}[b]{0.65\linewidth}
         \centering
         \includegraphics[width=\textwidth]{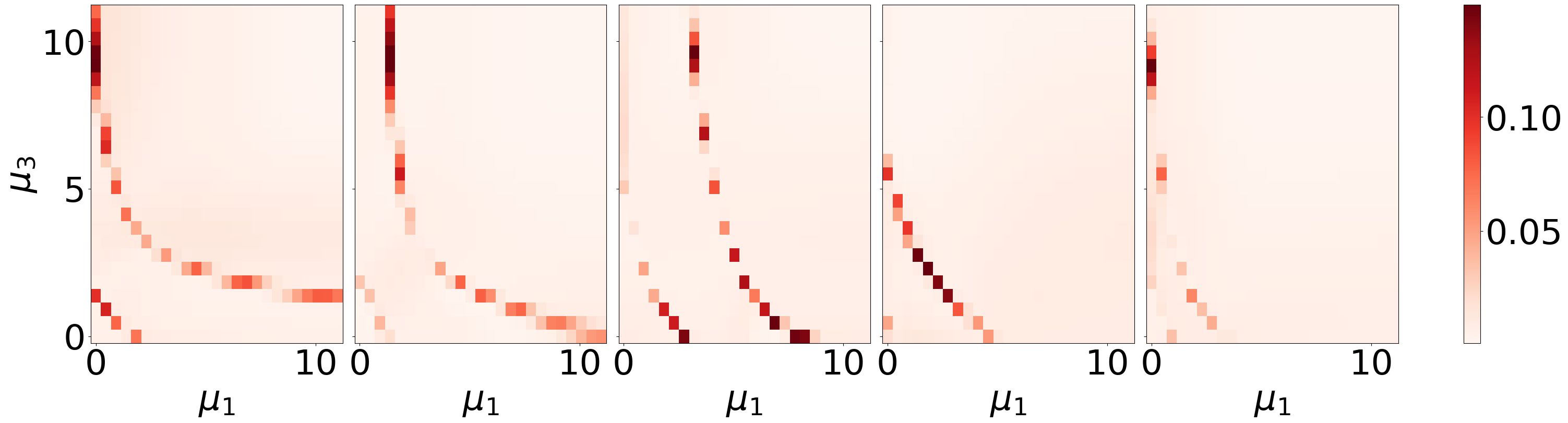}
         \caption{ }
     \end{subfigure}
     \begin{subfigure}[b]{0.65\linewidth}
         \centering
         \includegraphics[width=\textwidth]{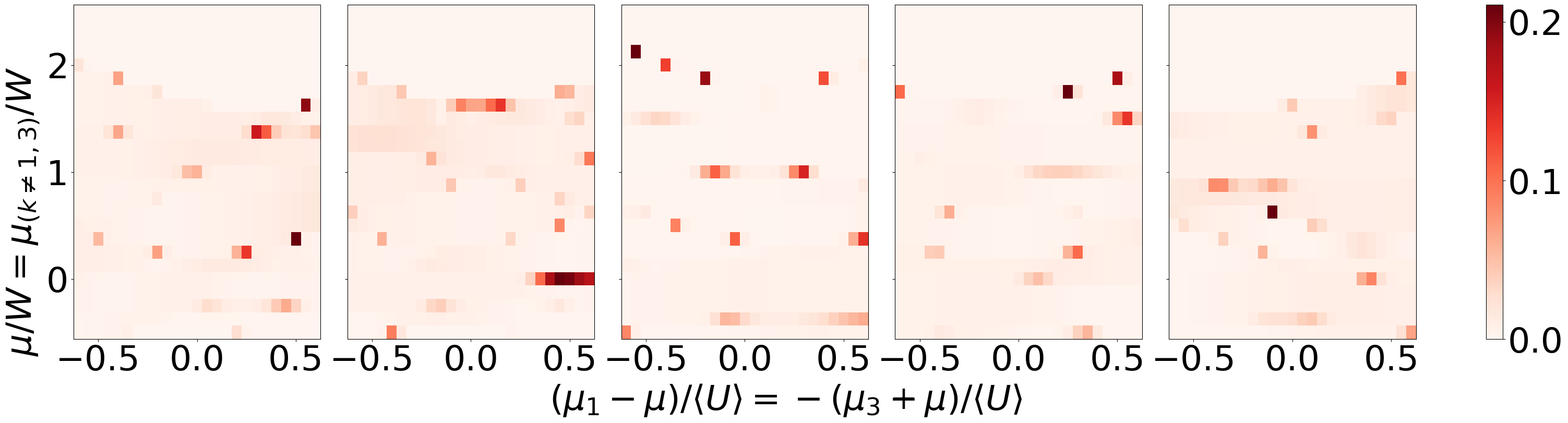}
         \caption{}
     \end{subfigure}
     \vspace{-0.2cm}
    \caption{{Difference in 4 quantum dot in charge stability diagrams for input and expected measurement outcomes from Hubbard model parameters for the prediction of disorder in all parameters. In particular $n_{error}=||\langle\vec{n}_{input}\rangle-\langle\vec{n}_{predicted}\rangle||$ is plotted in both plots where $\langle\vec{n}_{input}\rangle$ is the occupation expectation vector for all sites of the input data and $\langle\vec{n}_{predicted}\rangle$ is the expected occupation expectation vector given the prediction of the Hubbard model parameters. The columns of both plots correspond to the charge stability diagrams in Fig. \ref{fig:4AllParamFull} with the same column. The error-free model parameters were set to $U=4$, $t=1$ and $V_{i,i+1}=0.2$. (a) $n_{error}$ between input Hubbard parameters and predicted Hubbard parameters where $\mu_1$ and $\mu_3$ are independently varied. (b) $n_{error}$ between input Hubbard parameters and predicted Hubbard parameters where the chemical potential at each site is $\vec{\mu}=[\mu_1,\cdots,\mu_n]$ with our plot having axis $\mu_1=-\mu_3$ (rescaled by $\langle U \rangle=4$ and shifted by $\mu$) vs. $\mu_{k\neq 1,3}=\mu$ (rescaled by $W=\frac{1}{L}\left(\sum_i \langle U_i \rangle +\langle V_{i,i+1}\rangle\right)$ where $\langle U\rangle=4$ and $\langle V_{i,i+1}\rangle=0.2$ are the their disorder free values respectively)}}
         \label{fig:E4AllParamFull}
\end{figure}
\end{minipage}

\begin{minipage}{\linewidth} 
\begin{figure}[H]
     \centering
     \begin{subfigure}[b]{0.9\linewidth}
         \centering
         \includegraphics[width=\textwidth]{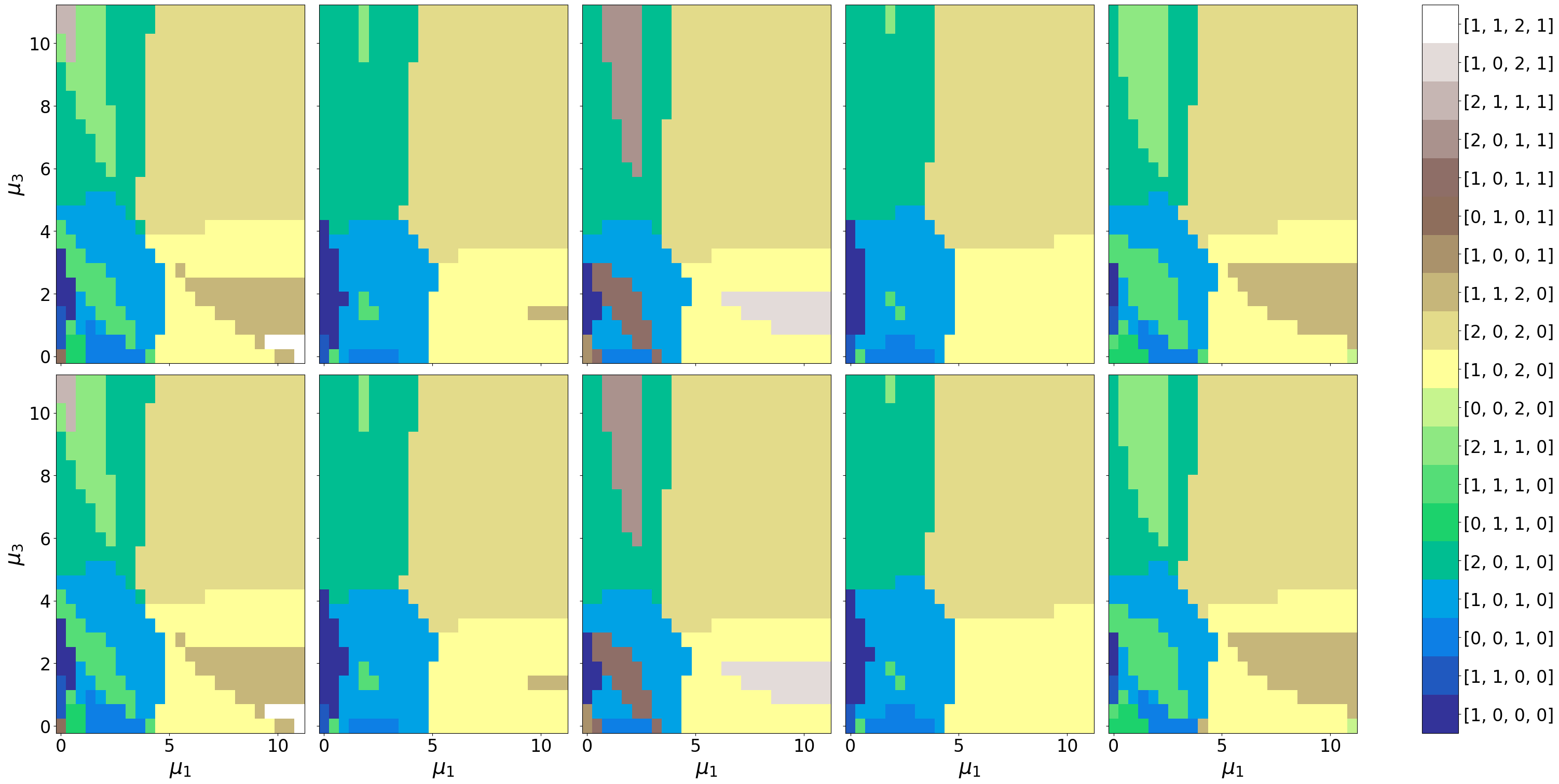}
         \caption{}
     \end{subfigure}
     \begin{subfigure}[b]{0.9\linewidth}
         \centering
         \includegraphics[width=\textwidth]{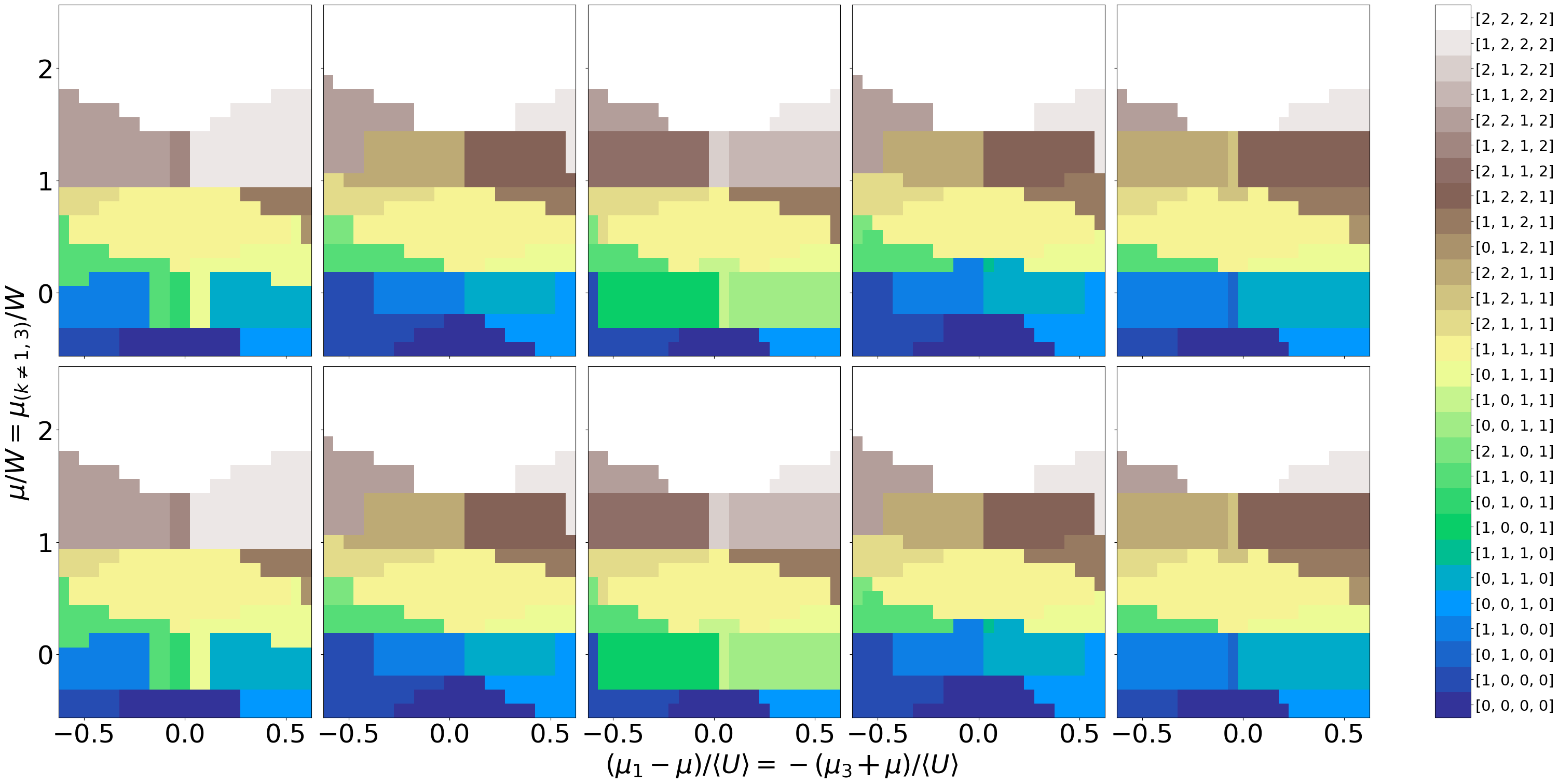}
         \caption{}
     \end{subfigure}
    \caption{{4 quantum dot charge stability diagrams for input and expected measurement outcomes from Hubbard model parameters for the prediction of only $\delta{\epsilon_i}$. The root mean squared error in the Hubbard model parameter was $R_{MS}({\delta \epsilon})=0.0029$ with an $R^2=0.99995$. In both plots the first row is the input charge stability diagrams, namely the most probable state for a chemical potential configuration, and the second row is the expected charge stability diagram given the prediction of the Hubbard model parameters. The 2D type CNN was used for this model. The error-free model parameters were set to $U=4$, $t=1$ and $V_{i,i+1}=0.2$. (a) Most probable state for input Hubbard parameters (1st row) and predicted Hubbard parameters (2nd row) where $\mu_1$ and $\mu_3$ are independently varied. (b) Most probable state for input Hubbard parameters (1st row) and predicted Hubbard parameters (2nd row) where the chemical potential at each site is $\vec{\mu}=[\mu_1,\cdots,\mu_n]$ with our plot having axis $\mu_1=-\mu_3$ (rescaled by $\langle U \rangle$ and shifted by $\mu$) vs. $\mu_{k\neq 1,3}=\mu$ (rescaled by $W=\frac{1}{L}\left(\sum_i \langle U_i \rangle +\langle V_{i,i+1}\rangle\right)$ where $\langle U\rangle=4$ and $\langle V_{i,i+1}\rangle=0.2$ are the their disorder free values respectively), this is similar to the stability diagrams fed into the neural network except the neural network only receives $\mu_{i}=-\mu_j$ between nearest neighbors $i=j+1$.}}
    \label{fig:4SingleParamEpsilonMu2}
\end{figure}
\end{minipage}

\begin{figure*}[h!]
     \centering
     \begin{subfigure}[b]{0.9\linewidth}
         \centering
         \includegraphics[width=\textwidth]{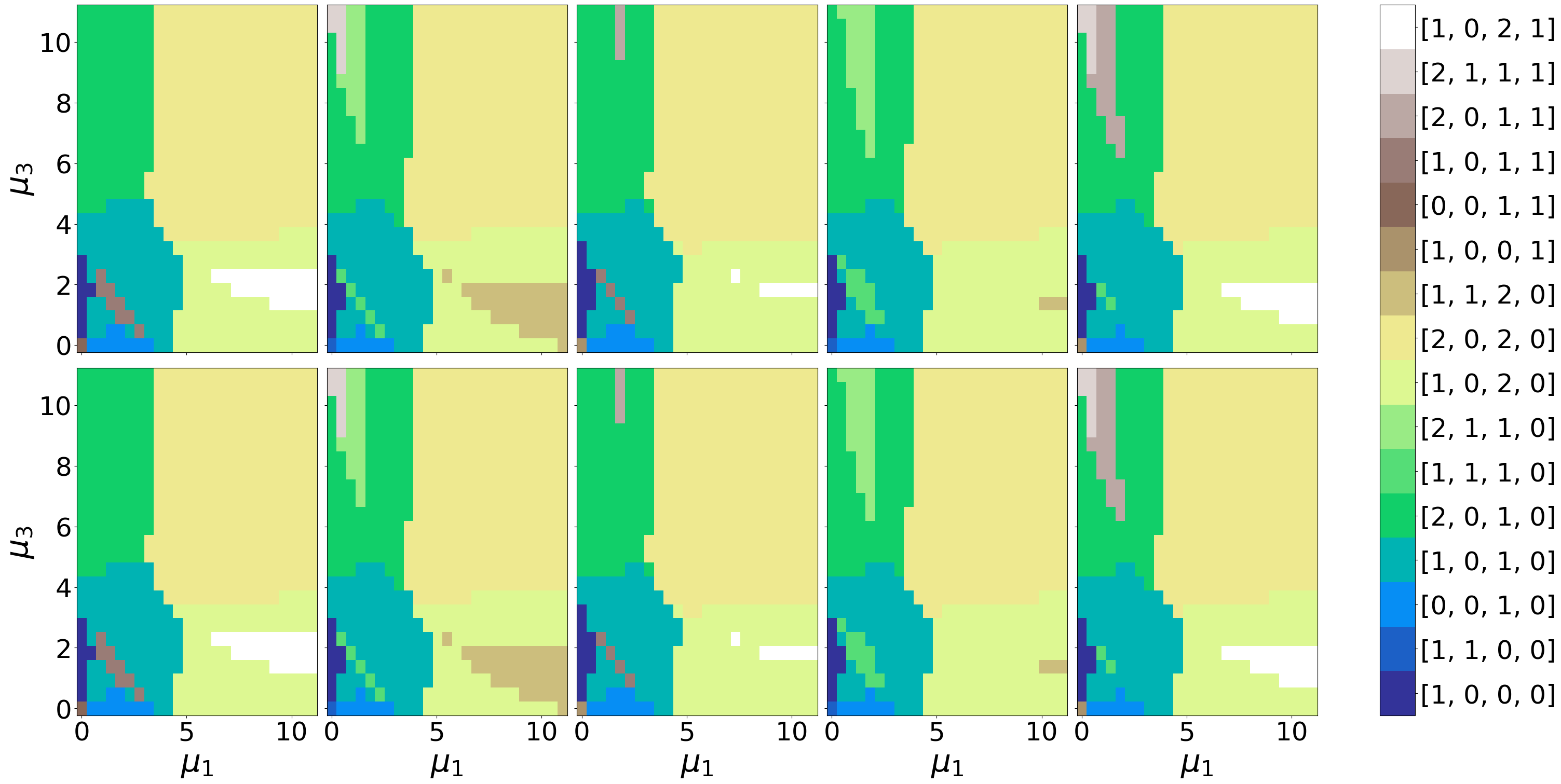}
         \caption{}
     \end{subfigure}
     \begin{subfigure}[b]{0.9\linewidth}
         \centering
         \includegraphics[width=\textwidth]{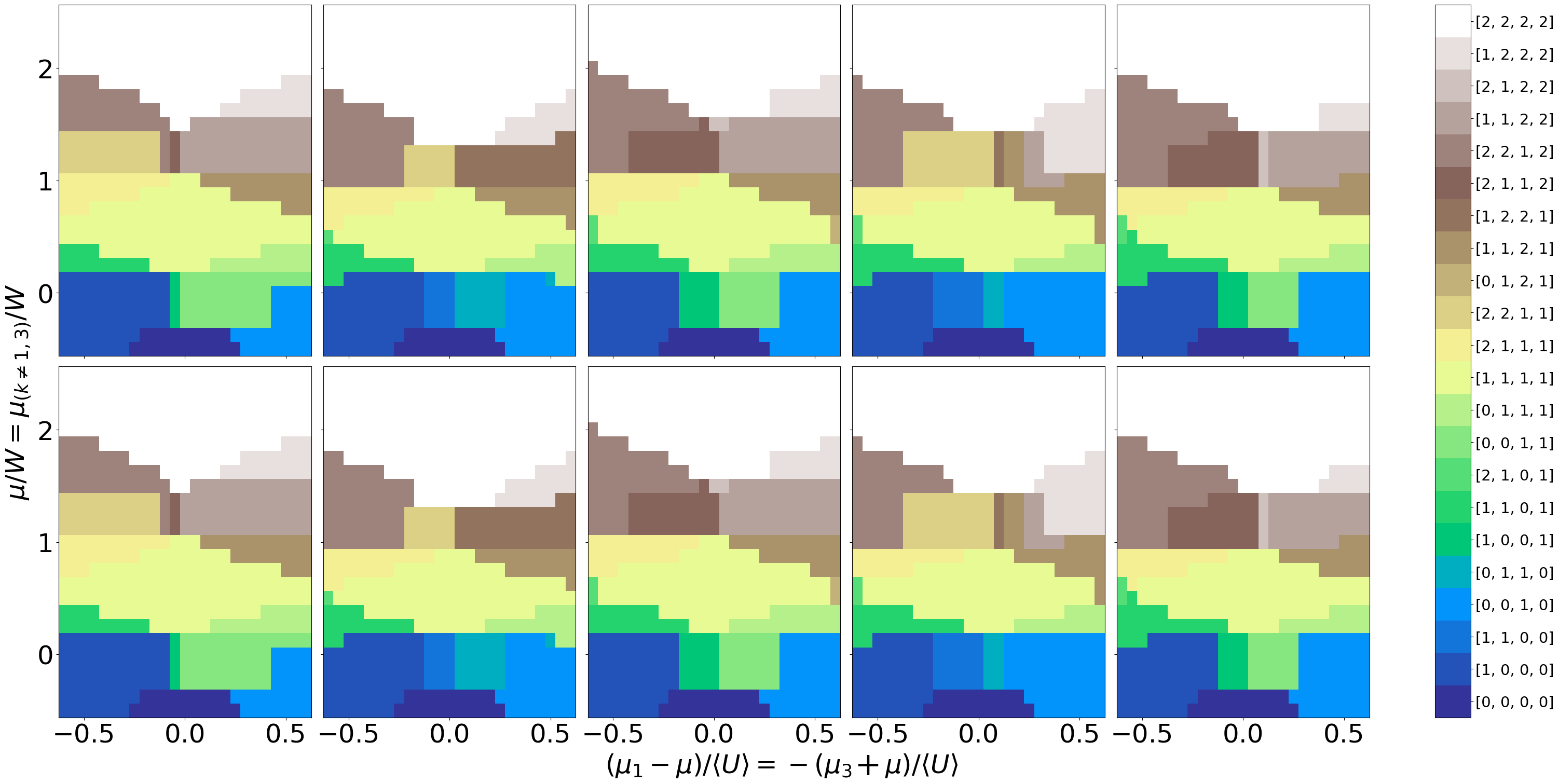}
         \caption{}
     \end{subfigure}

    \caption{4 quantum dot charge stability diagrams for input and expected from Hubbard model parameters for the prediction of only $\delta V_{ij}$. The root mean squared error in the Hubbard model $\delta V_{ij}$ parameter was $R_{MS}({\delta V_{ij}})=$ with an $R^2=0.99995$. In both plots the first row is the input charge stability diagrams, namely the most probable state for a chemical potential configuration, and the second row is the expected charge stability diagram given the prediction of the Hubbard model parameters. The error-free model parameters were set to $U=4$, $t=1$ and $V_{i,i+1}=0.2$. The columns of these two subplots correspond to the same representative samples. (a) Most probable state for input Hubbard parameters (1st row) and predicted Hubbard parameters (2nd row) where $\mu_1$ and $\mu_3$ are independently varied. (b) Most probable state for input Hubbard parameters (1st row) and predicted Hubbard parameters (2nd row) where the chemical potential at each site is $\vec{\mu}=[\mu_1,\cdots,\mu_n]$ with our plot having axis $\mu$ (rescaled by $\langle U \rangle$ and shifted by $\mu$) vs. $\mu_{k\neq 1,3}=\mu$ (rescaled by $W=\frac{1}{L}\left(\sum_i \langle U_i \rangle +\langle V_{i,i+1}\rangle\right)$ where $\langle U\rangle=4$ and $\langle V_{i,i+1}\rangle=0.2$ are the their disorder free values respectively), this is similar to the stability diagrams fed into the neural network except the neural network only receives $\mu_{i}=-\mu_j$ between nearest neighbors $i=j+1$.}     \label{fig:4SingleParamVij}
\end{figure*}

\begin{figure*}[h!]
     \centering
     \begin{subfigure}[b]{0.9\linewidth}
         \centering
         \includegraphics[width=\textwidth]{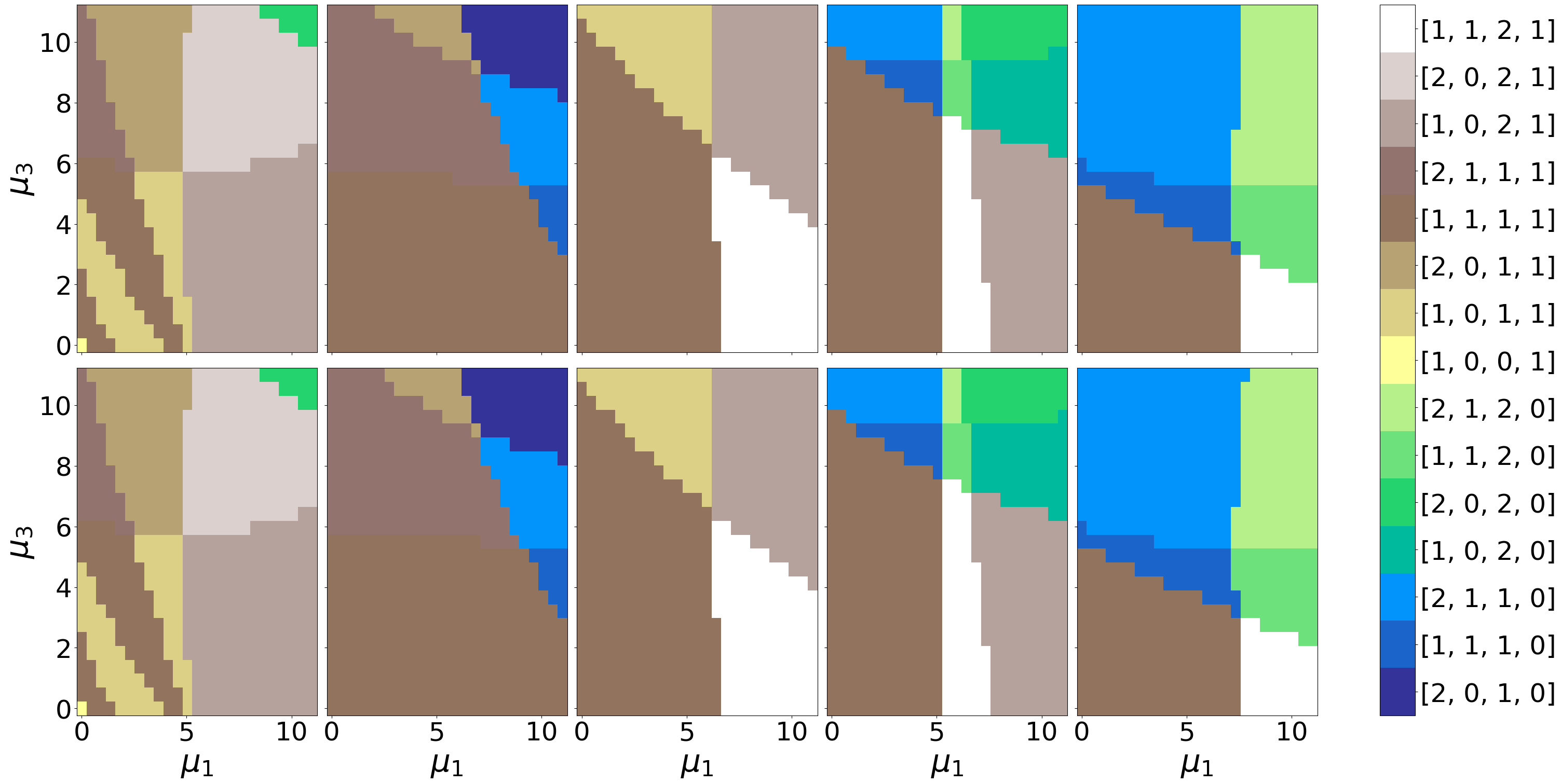}
         \caption{}
     \end{subfigure}
     \begin{subfigure}[b]{0.9\linewidth}
         \centering
         \includegraphics[width=\textwidth]{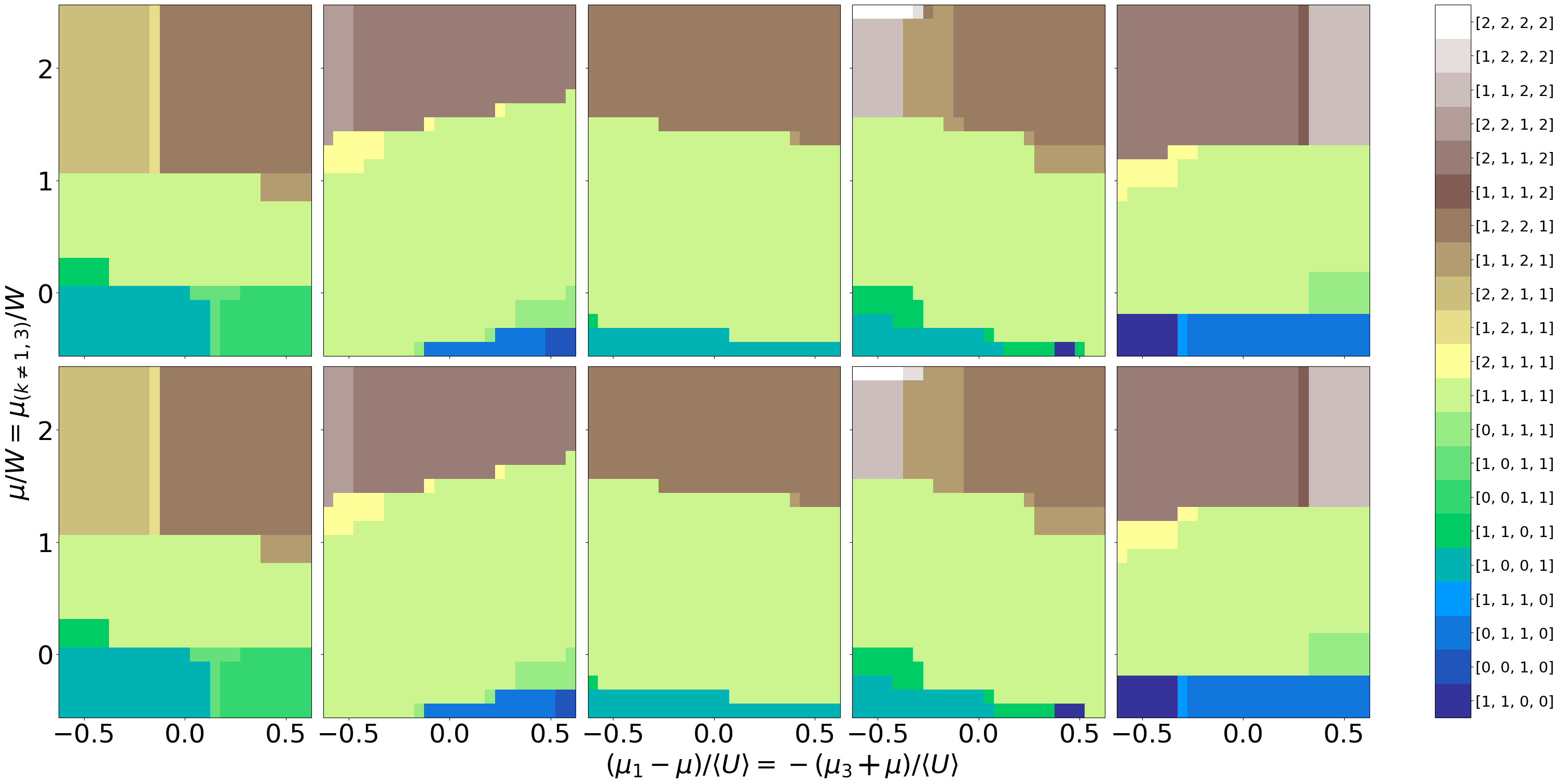}
         \caption{}
     \end{subfigure}

    \caption{4 quantum dot charge stability diagrams for input and expected from Hubbard model parameters for the prediction of only $\delta t_{ij}$. The root mean squared error in the Hubbard model parameter was $R_{MS}({\delta t_{ij}})=0.0243$ with an $R^2=0.99997$. In both plots the first row is the input charge stability diagrams, namely the most probable state for a chemical potential configuration, and the second row is the expected charge stability diagram given the prediction of the Hubbard model parameters. The error-free model parameters were set to $U=4$, $t=1$ and $V_{i,i+1}=0.2$. The columns of these two subplots correspond to the same representative samples. (a) Most probable state for input Hubbard parameters (1st row) and predicted Hubbard parameters (2nd row) where $\mu_1$ and $\mu_3$ are independently varied.  (b) Most probable state for input Hubbard parameters (1st row) and predicted Hubbard parameters (2nd row) where the chemical potential at each site is $\vec{\mu}=[\mu_1,\cdots,\mu_n]$ with our plot having axis $\mu_1=-\mu_3$ (rescaled by $\langle U \rangle$ and shifted by $\mu$) vs. $\mu_{k\neq 1,3}=\mu$ (rescaled by $W=\frac{1}{L}\left(\sum_i \langle U_i \rangle +\langle V_{i,i+1}\rangle\right)$ where $\langle U\rangle=4$ and $\langle V_{i,i+1}\rangle=0.2$ are the their disorder free values respectively ), this is similar to the stability diagrams fed into the neural network except the neural network only receives $\mu_{i}=-\mu_j$ between nearest neighbors $i=j+1$.}     \label{fig:4SingleParamdt}
\end{figure*}

\begin{figure*}[h!]
     \centering
     \begin{subfigure}[b]{0.9\linewidth}
         \centering
         \includegraphics[width=\textwidth]{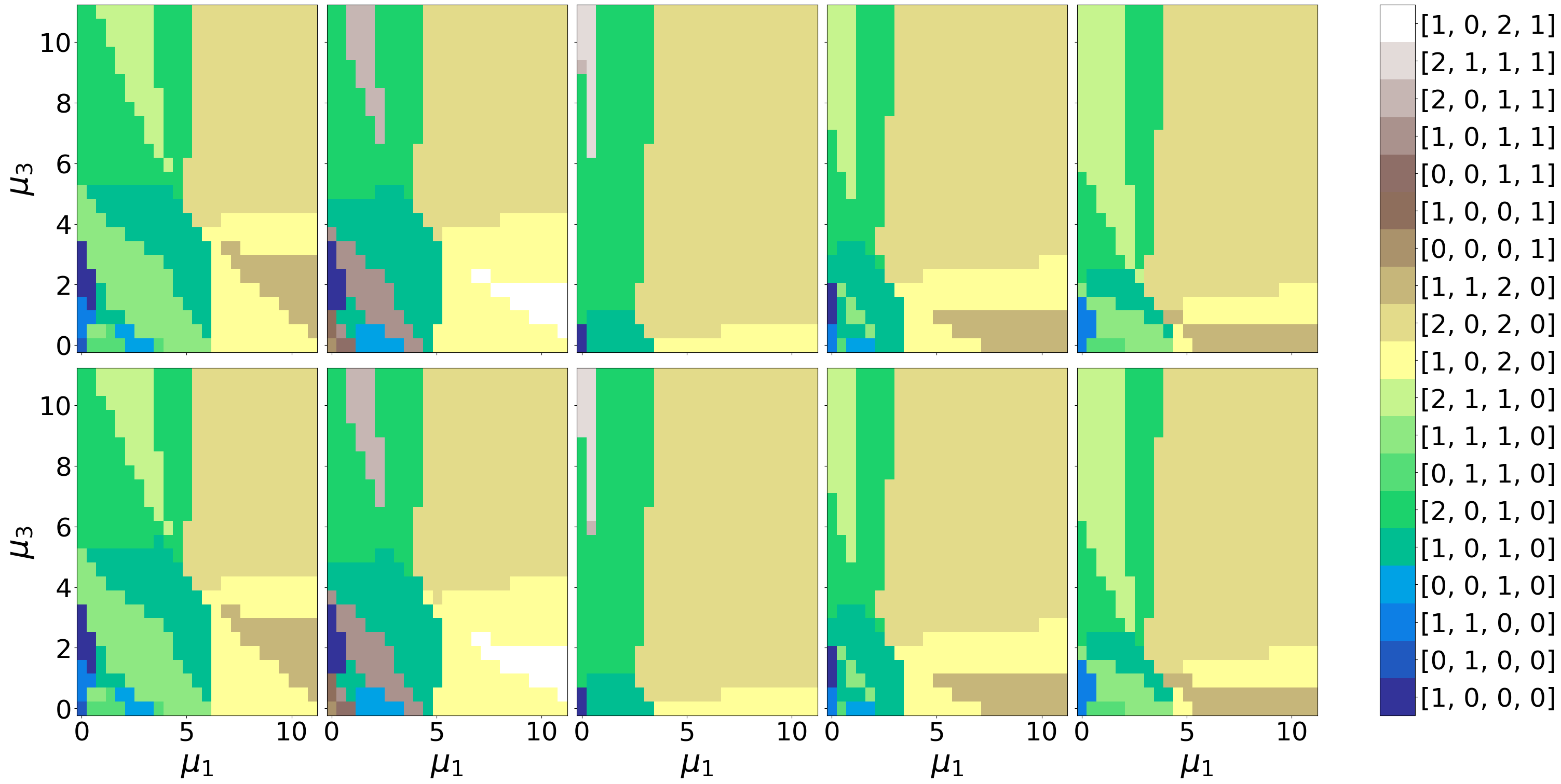}
         \caption{}
     \end{subfigure}
     \begin{subfigure}[b]{0.9\linewidth}
         \centering
         \includegraphics[width=\textwidth]{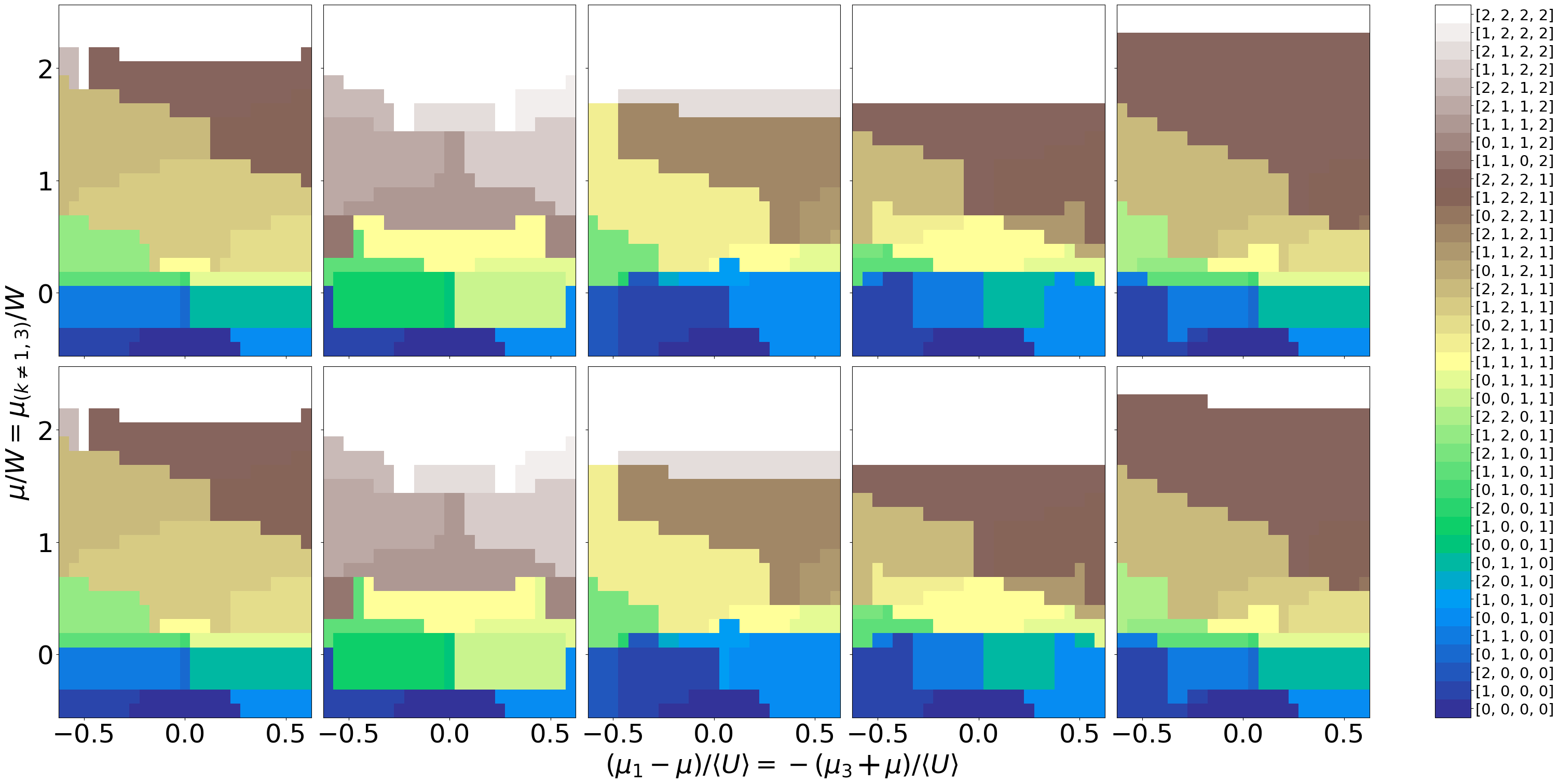}
         \caption{}
     \end{subfigure}

    \caption{4 quantum dot charge stability diagrams for input and expected from Hubbard model parameters for the prediction of only $\delta U_i$. The root mean squared error in the Hubbard model $\delta U_i$ parameter was $R_{MS}({\delta U_{i}})=0.1005$ with an $R^2=0.9987$. In both plots the first row is the input charge stability diagrams, namely the most probable state for a chemical potential configuration, and the second row is the expected charge stability diagram given the prediction of the Hubbard model parameters. The error-free model parameters were set to $U=4$, $t=1$ and $V_{i,i+1}=0.2$. The columns of these two subplots correspond to the same representative samples. (a) Most probable state for input Hubbard parameters (1st row) and predicted Hubbard parameters (2nd row) where $\mu_1$ and $\mu_3$ are independently varied. (b) Most probable state for input Hubbard parameters (1st row) and predicted Hubbard parameters (2nd row) where the chemical potential at each site is $\vec{\mu}=[\mu_1,\cdots,\mu_n]$ with our plot having axis $\mu_1=-\mu_3$ (rescaled by $\langle U \rangle$ and shifted by $\mu$) vs. $\mu_{k\neq 1,3}=\mu$ (rescaled by $W=\frac{1}{L}\left(\sum_i \langle U_i \rangle +\langle V_{i,i+1}\rangle\right)$ where $\langle U\rangle=4$ and $\langle V_{i,i+1}\rangle=0.2$ are the their disorder free values respectively), this is similar to the stability diagrams fed into the neural network except the neural network only receives $\mu_{i}=-\mu_j$ between nearest neighbors $i=j+1$.}     \label{fig:4SingleParamUi}
\end{figure*}

\begin{minipage}{\linewidth} 
\begin{figure}[H]
     \centering
     \begin{subfigure}[b]{0.9\linewidth}
         \centering
         \includegraphics[width=\textwidth]{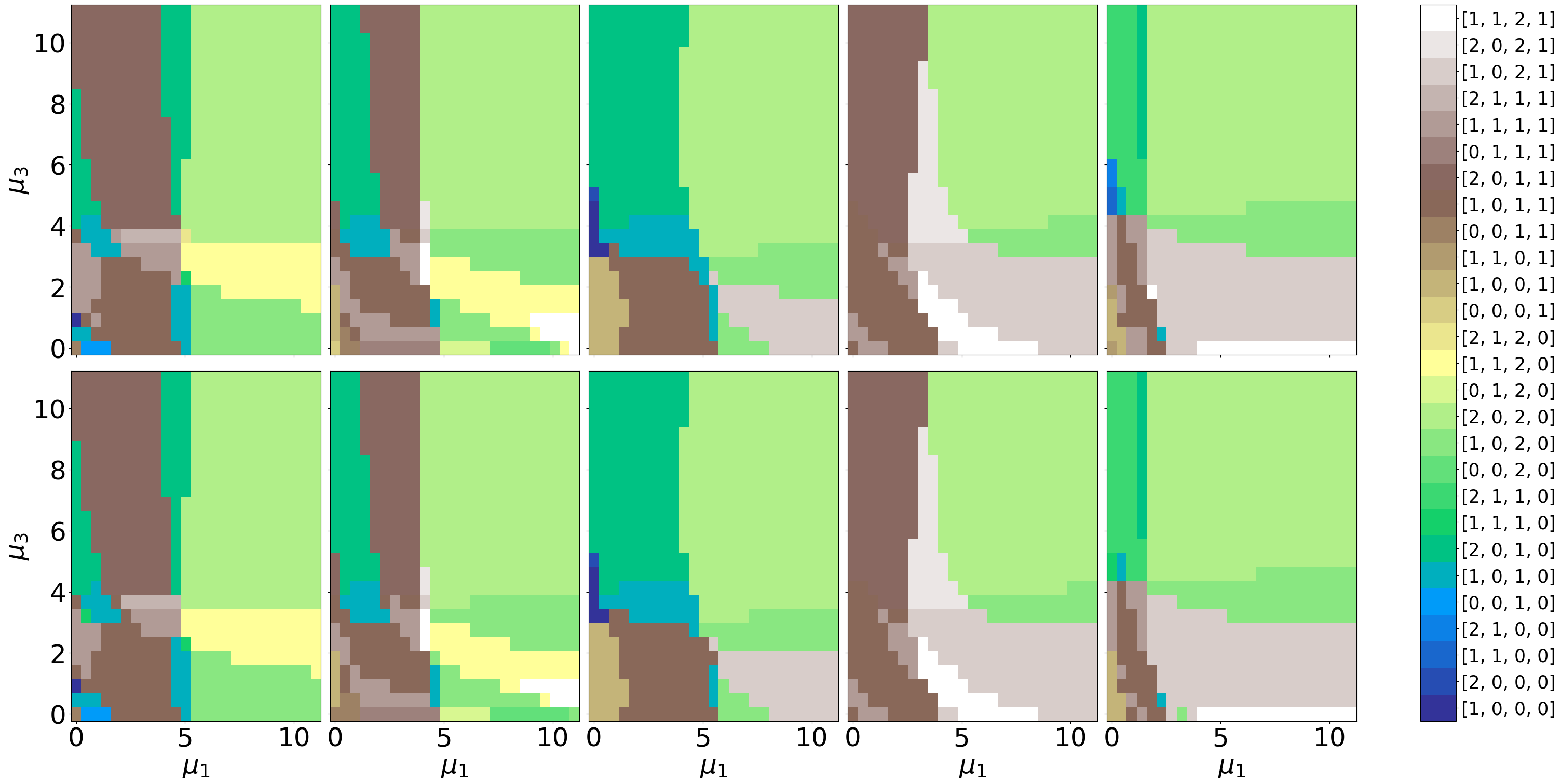}
         \caption{}
     \end{subfigure}
     \begin{subfigure}[b]{0.9\linewidth}
         \centering
         \includegraphics[width=\textwidth]{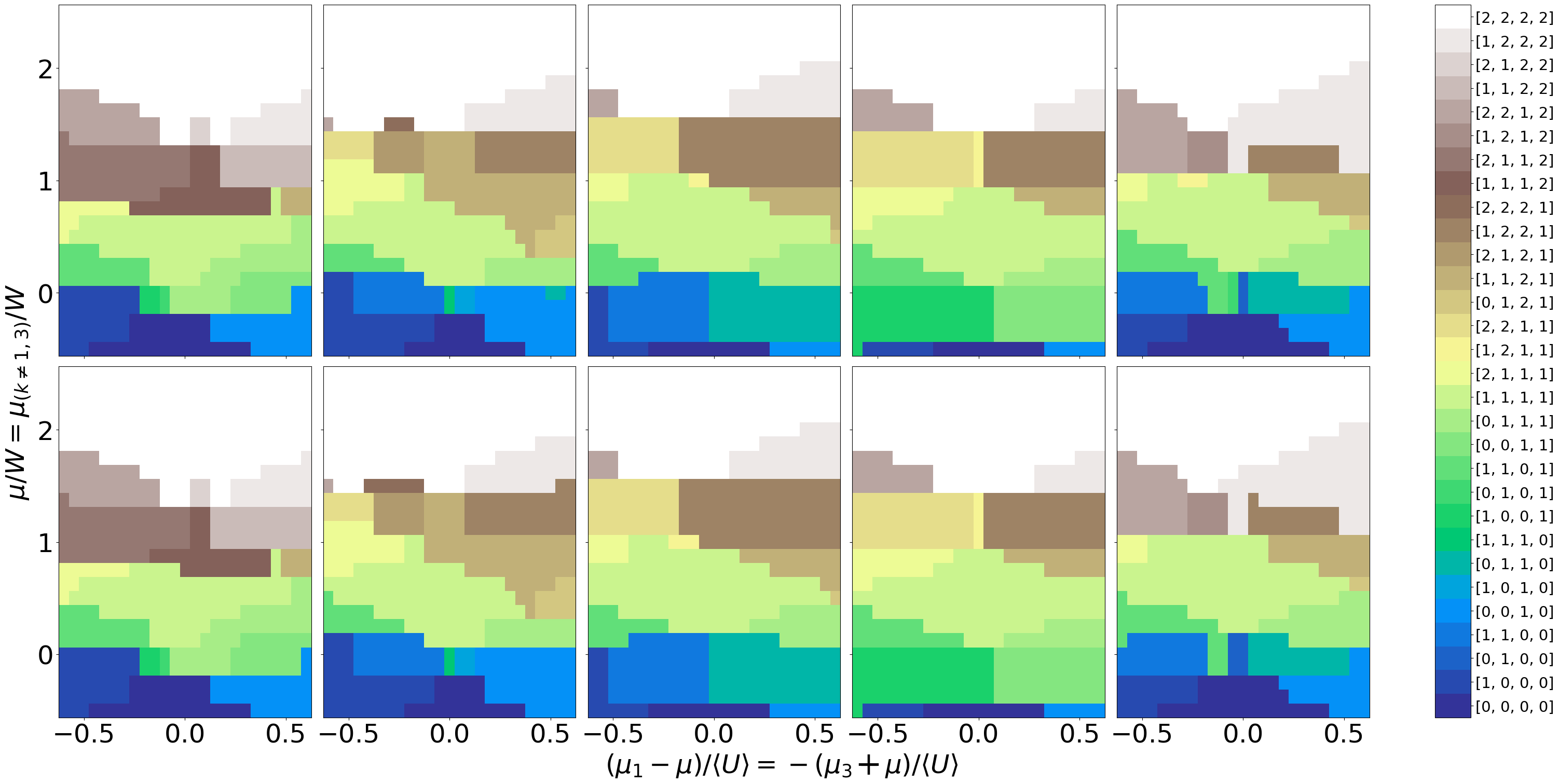}
         \caption{}
     \end{subfigure}

    \caption{4 quantum dot charge stability diagrams for input and expected measurement outcomes from Hubbard model parameters for the prediction of disorder in all parameters. The root mean squared error in the Hubbard model parameters was $R_{MS}({\delta \epsilon_{i}})=0.0280$, $R_{MS}(\delta V_{i,j})=0.0205$, $R_{MS}(\delta t_{i,j})=0.0193$, $R_{MS}(\delta U_{i})=0.0512$ with an $R^2=0.9961$. In both plots the first row is the input charge stability diagrams, namely the most probable state for a chemical potential configuration, and the second row is the expected charge stability diagram given the prediction of the Hubbard model parameters. (a) Most probable state for input Hubbard parameters (1st row) and predicted Hubbard parameters (2nd row) where $\mu_1$ and $\mu_3$ are independently varied. (b) Most probable state for input Hubbard parameters (1st row) and predicted Hubbard parameters (2nd row) where the chemical potential at each site is $\vec{\mu}=[\mu_1,\cdots,\mu_n]$ with our plot having axis $\mu_1=-\mu_3$ (rescaled by $\langle U \rangle$ and shifted by $\mu$) vs.  $\mu_{k\neq 1,3}=\mu$ (rescaled by $W=\frac{1}{L}\left(\sum_i \langle U_i \rangle +\langle V_{i,i+1}\rangle\right)$ where $\langle U\rangle=4$ and $\langle V_{i,i+1}\rangle=0.2$ are the their disorder free values respectively ), this is similar to the stability diagrams fed into the neural network except the neural network only receives $\mu_{i}=-\mu_j$ between nearest neighbors $i=j+1$.
    }     \label{fig:4AllParamFull}
\end{figure}
\end{minipage}

\begin{figure*}[h!]
     \centering
     \begin{subfigure}[b]{0.9\linewidth}
         \centering
         \includegraphics[width=\textwidth]{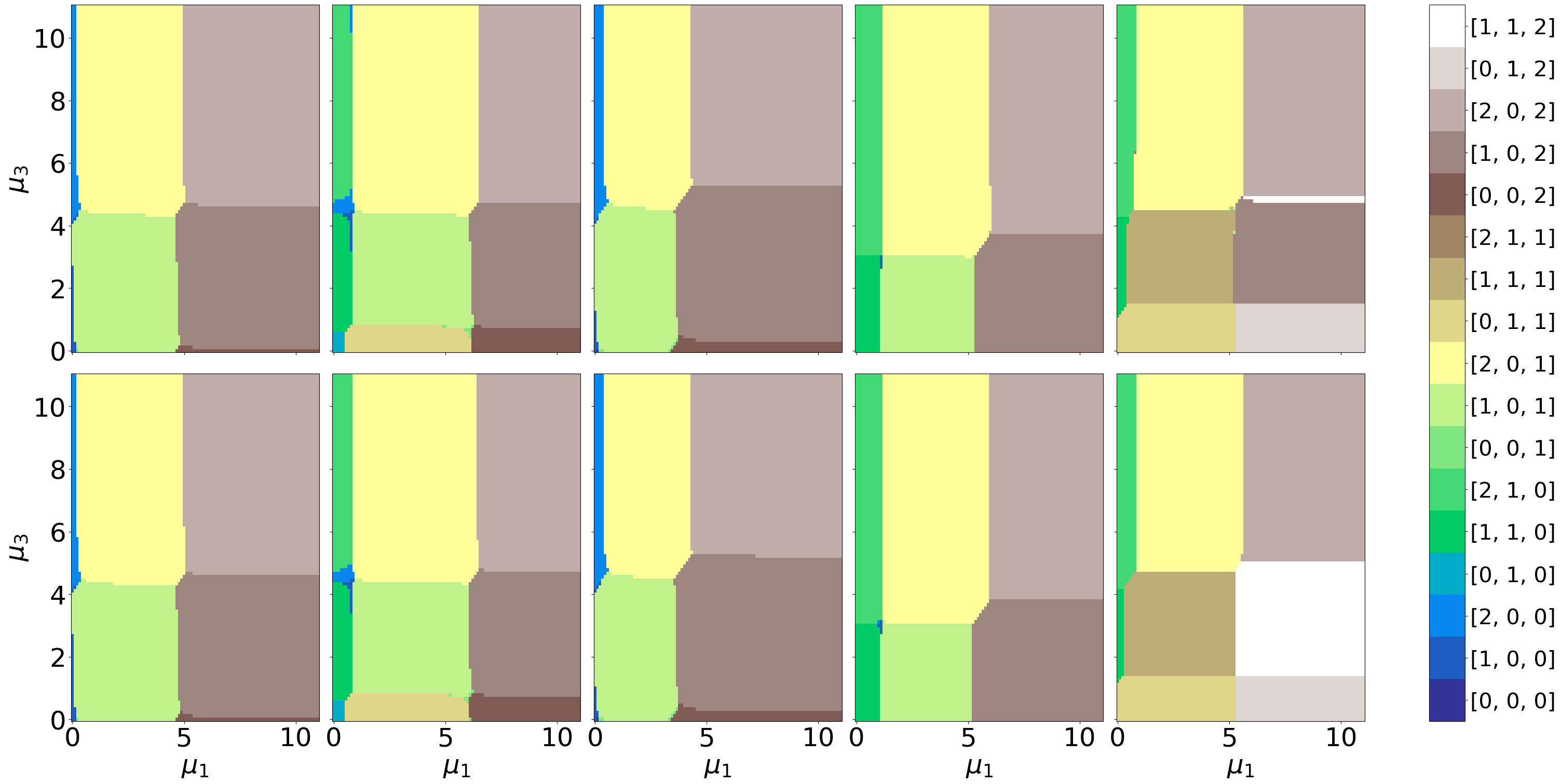}
         \caption{}
     \end{subfigure}
     \begin{subfigure}[b]{0.9\linewidth}
         \centering
         \includegraphics[width=\textwidth]{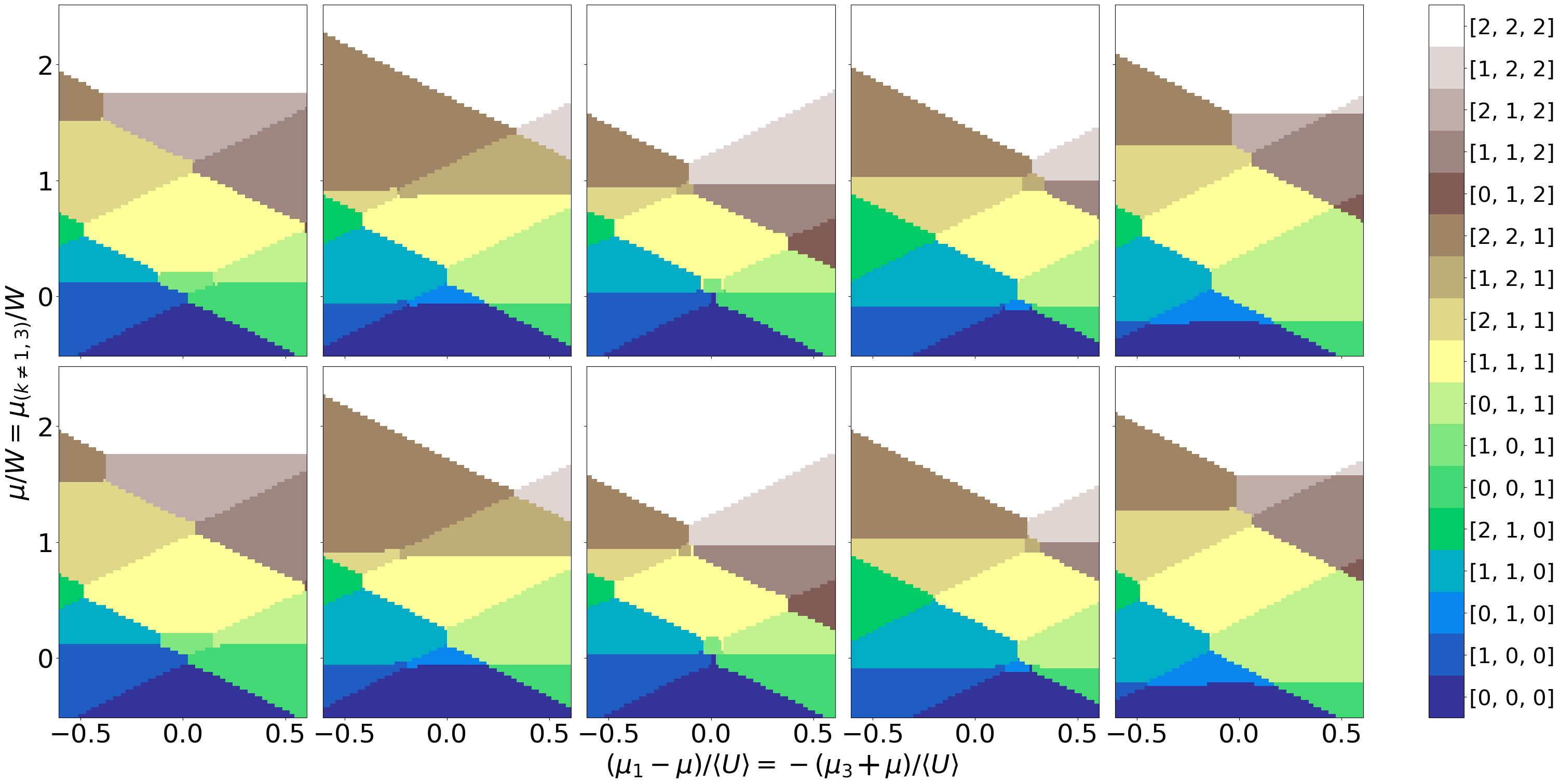}
         \caption{}
     \end{subfigure}

    \caption{3 quantum dot charge stability diagrams for input and expected measurement outcomes from Hubbard model parameters for the prediction of disorder in all parameters with higher resolution, standard scaled, and reduced range $\tilde{t} \in [0.01,0.25]$. The root mean squared error in the Hubbard model parameters was $R_{MS}({\delta \epsilon_{i}})=0.0110$, $R_{MS}({\delta V_{i,j}})=0.0113$, $R_{MS}({\delta t_{i,j}})=0.0045$, $R_{MS}({\delta U_{i}})=0.0347$ with an $R^2=0.9972$. In both plots the first row is the input charge stability diagrams, namely the most probable state for a chemical potential configuration, and the second row is the expected charge stability diagram given the prediction of the Hubbard model parameters. (a) Most probable state for input Hubbard parameters (1st row) and predicted Hubbard parameters (2nd row) where $\mu_1$ and $\mu_3$ are independently varied. (b) Most probable state for input Hubbard parameters (1st row) and predicted Hubbard parameters (2nd row) where the chemical potential at each site is $\vec{\mu}=[\mu_1,\cdots,\mu_n]$ with our plot having axis $\mu_1=-\mu_3$ (rescaled by $\langle U \rangle$ and shifted by $\mu$) vs. $\mu_{k\neq 1,3}=\mu$ (rescaled by $W=\frac{1}{L}\left(\sum_i \langle U_i \rangle +\langle V_{i,i+1}\rangle\right)$ where $\langle U\rangle=4$ and $\langle V_{i,i+1}\rangle=0.2$ are the their disorder free values respectively), this is the data input into the machine learning model. 
    }\label{fig:3AllParamFullSSHighRes}
\end{figure*}

\end{document}